# Title: Universality of clone dynamics during tissue development

**Authors:** Steffen Rulands[1-5,*], Fabienne Lescroart[6], Samira Chabab[6], Christopher J. Hindley[1,2], Nicole Prior[2], Magdalena K. Sznurkowska[2,7], Meritxell Huch[2,3], Anna Philpott[2,7], Cedric Blanpain[6] and Benjamin D. Simons[1-3,*]

**Affiliations:**

[1]Cavendish Laboratory, Department of Physics, JJ Thomson Avenue, University of Cambridge, Cambridge CB3 0HE, UK.

[2]The Wellcome Trust/Cancer Research UK Gurdon Institute, University of Cambridge, Tennis Court Road, Cambridge CB2 1QN, UK.

[3]Wellcome Trust Centre for Stem Cell Research, University of Cambridge, Tennis Court Road, Cambridge CB2 1QR, UK.

[4]Max Planck Institute for the Physics of Complex Systems, Noethnitzer Str. 38, 01187 Dresden Germany

[5]Center for Systems Biology Dresden, Pfotenhauer Str. 108, 01307 Dresden, Germany

[6]Université Libre de Bruxelles, Laboratory of Stem Cells and Cancer, Brussels B-1070, Belgium.

[7]Department of Oncology, University of Cambridge, Hutchison/MRC Research Centre, Hills Road, Cambridge CB2 0XZ, UK.

* Corresponding authors

**The emergence of complex organs is driven by the coordinated proliferation, migration and differentiation of precursor cells. The fate behaviour of these cells is reflected in the time evolution their progeny, termed clones, which serve as a key experimental observable. In adult tissues, where cell dynamics is constrained by the condition of homeostasis, clonal tracing studies based on transgenic animal models have advanced our understanding of cell fate behaviour and its dysregulation in disease (*1*, *2*). But what can be learned from clonal dynamics in development, where the spatial cohesiveness of clones is impaired by tissue deformations during tissue growth? Drawing on the results of clonal tracing studies, we show that, despite the complexity of organ development, clonal dynamics may converge to a critical state characterized by universal scaling behaviour of clone sizes. By mapping clonal dynamics onto a generalization of the classical theory of aerosols, we elucidate the origin and range of scaling behaviours and show how the identification of universal scaling dependences may allow lineage-specific information to be distilled from experiments. Our study shows the emergence of core concepts of statistical physics in an unexpected context, identifying cellular systems as a laboratory to study non-equilibrium statistical physics.**

Biological systems, being highly structured and dynamic, function far from thermal equilibrium. This is particularly evident in embryonic development where, through large-scale cellular self-organisation, highly complex structures emerge from a group of genetically identical, pluripotent stem cells. To achieve the stereotypic ordering of organs and tissues, the fate of embryonic stem cells and their progeny must be tightly-regulated, such that the correct number and type of cells is generated at the right time and place during development. Mechanisms regulating such cell fate decisions are at the center of research in stem cell and developmental biology (*3*). Efforts to resolve the mechanisms that regulate cell fate behaviour place emphasis on emerging technologies, including single-cell genomics and genome editing methods, which provide detailed information on the subcellular and cellular processes. However, by focusing on gene regulatory programmes, such approaches often fail to engage with how collective cell behaviour, and the formation of functioning organs, emerges from the network of complex interactions at the molecular scale.

To understand how complexity at the microscopic scale translates into coherent collective behaviour at the macro-scale, statistical physics provides a useful theoretical framework. For critical systems, where fluctuations are scale-invariant, successive coarse-graining can yield effective theories describing macroscopic behavior. In such systems, different "microscopic" systems can give rise to indistinguishable macroscopic behavior – a concept known as universality. As a reflection of scale invariance, statistical correlations, such as size distributions, obtain simple scaling forms, which depend only on one or few dimensionless composite variables. But, given the complexity of embryonic development, can such concepts be applied to study cellular behaviour?

At the cellular scale, the patterns of cell fate decisions during embryonic development are reflected in the time-evolution of individual developmental precursors cells and their progeny, which together constitute a clone. While the dynamics of individual clones maybe complex, subject both to intrinsic and extrinsic influences, statistical ensembles of clones may provide robust (predictive) information about the relationship between different cell types and mechanisms regulating cellular behaviour. In mammals, where live-imaging of developing embryonic organs is typically infeasible, efforts to resolve clonal dynamics have relied on cell lineage tracing studies using transgenic animal models (*1*). In this approach, the activation of a reporter gene allows individual cells to be marked with a fluorescent reporter. As a genetic mark, this label is then inherited by all progeny of a marked cell, and allows clone sizes and cell compositions to be recorded at specific times post-labelling (Figure 1A). Lineage tracing studies therefore provide a "two-time" measure of clonal dynamics in the living embryo. In adult tissues, where cell dynamics is heavily constrained by the steady state condition of homeostasis, efforts to resolve cell fate behaviour from clonal tracing studies have drawn successfully upon concepts from statistical physics and mathematics (*4–6*). However, in developing tissues, the interpretation of these experiments is complicated by the fact that clonal dynamics is, in principle, less constrained. Moreover, due to large-scale cellular rearrangements as well as stochastic forces from surrounding tissues, labelled clones may fragment into disconnected clusters, or they merge and form larger compounds of labelled cells (Figure 1B-F).

Here, by establishing a formal mapping between clonal dynamics and a generalization of the theory of aerosols, we show that, during embryonic development, clonal dynamics converges to a critical state, giving rise to universal scaling behaviour of the size distributions of labelled clusters. Further, we explore how understanding the origins of scaling and universality can form the quantitative basis for recovering information on cell fate behaviour during development. We thus find the emergence of core concepts of statistical physics in the unexpected context of embryonic development. As well as being of interest in the study of tissue development, these findings have important implications for the study of tissue regeneration and tumour growth.

To develop this programme, we begin with an example of clonal evolution during the development of mouse heart. The gene Mesp1 is transiently expressed between embryonic day (E)6.5 and E7.5 in mice in the earliest precursor cells of the heart (*7–9*). Quantitative analysis of hearts labelled at low density (1-2 clones per heart) have established the temporal progression in differentiation and proliferative capacity of these precursors (*8*, *9*). However, with just 1 or 2 clones per embryo, and inherent variability in the efficiency of labelling, low-density labelling is highly inefficient in probing evolutionary processes during development. By contrast, at high (mosaic) labelling density, each embryo provides a potentially rich dataset. The situation is exemplified in Figure 1E, which shows mouse hearts at E12.5 and postnatal day P1 after mosaic labelling between E6.5 and E7.5 using the multicolour *Mesp1-Cre/Rosa-Confetti* reporter construct (with 50% of the cardiac surface being fluorescently labelled with each of three colours, cyan, yellow and red, roughly equally represented). However, at this density of labelling, a single contiguous cluster of labelled cells can be derived from the chance fusion of two or more independent clones induced with the same colour (*10*). Given that clone sizes are not constrained by tissue size, and the ambiguity arising from clone merger and fragmentation, to what extent can information on cell fate behaviour be recovered?

To address this question, we quantified the surface area (SA) covered by each cluster in a given heart compartment at different developmental time points. From the SAs, we then determined their distributions in each heart region (Fig. 1F). Although cardiac development involves complex cell fate decisions, with regional and temporal variations in proliferation (*11*, *12*), we found that the resulting cluster size distribution was remarkably conserved: After rescaling the SA of each cluster by the ensemble average for each compartment at a given time point, the resulting rescaled size distributions perfectly overlapped (Fig. 1G,H). This result implies that, despite the complex and variable histories, the resulting SA distribution is fully characterized by the average alone, the defining property of scaling. Formally, the frequency $f(x,t)dx$ of a cluster with a SA between $x$ and $x + dx$ at time $t$ post-labelling acquires the statistical scaling form, $f(x,t) = \phi(x/\langle x(t)\rangle)$, where $\phi$ denotes the scaling function.

The simplicity of the cluster size distribution that is reflected in scaling behaviour suggests that its origin may not rely on details of the morphogenic programme in heart. Rather, to uncover its origin, we began by considering the simplest set of processes that could determine cluster size: First, as labelled cells divide, clusters may grow at a rate proportional to their size. Second, in expanding tissues, clones may fragment into disconnected clusters as cells disperse or the tissue deforms. If the rate of growth and fragmentation increase in proportion to cluster size, the SA distribution would be predicted to become stationary. However, although clonal tracing studies indicate that growth and fragmentation occur on a similar time scale during the early phase of heart development (E6.5 and E12.5) (*9*), average cluster sizes at E12.5 and P1 differ by a factor of 2.7, showing that steady-state is not reached. More importantly, such a simple line of argument neglects the possibility that clusters of the same colour can merge into larger, cohesively labelled regions. Yet the number of clusters varies only marginally between E12.5 and P1 (*9*), indicating that merger and fragmentation could be equally abundant.

To resolve the origin of scaling, it is instructive to leave temporarily the realm of biology and consider the growth dynamics of "inanimate" compounds. Indeed, processes involving merger and fragmentation occur in multiple contexts in physics, including the nucleation of nano-crystals, amyloid fibrils, polymerisation, endocytosis and the dynamics of aerosols (*13–16*). In common with clonal evolution in tissues, droplets in aerosols may merge (coagulate) or they may fragment (Fig. 2A). By analogy with clonal growth due to cell division, droplets may also expand by condensation of free molecules, while cell loss due to death or migration out of the imaging window is mirrored in the evaporation and shrinking of droplet sizes. Finally, by analogy with the migration of cells into the field of view, new droplets may nucleate from free molecules. Through this correspondence, can the statistical physics of aerosols provide insight into the dynamics of cell clusters in tissues and the emergence of scaling?

The distribution of cluster sizes, $f(x,t)$, is the result of different sources of variability including merger, fragmentation, cell division and loss. Formally, the time evolution of the cluster size distribution can be cast (symbolically) as a sum of operators that describe the effect of these contributions on the time evolution,

$$\partial_t f(x,t) = L_{\text{growth}}[f(x,t)] + \varphi\, L_{\text{fragmentation}}[f(x,t)] + \mu\, L_{\text{merger}}[f(x,t)] + \cdots,$$

where the parameters, $\varphi$, $\mu$, etc. characterize the relative strength of these processes against that of growth (for details, see Supplementary Theory). To investigate the origin of scaling, we questioned what determines the long-term, large-scale dependence of the cluster size distribution. In statistical physics this question is typically answered by successively coarse-graining the dynamics and monitoring changes in the relative contributions of different processes. Under this renormalization, when a cell divides, cluster sizes are rescaled by the resulting increase in tissue size, $x \to x/(1+\delta X) \equiv \rho$. Simultaneously, time is rescaled in such way that the total rate of merging and fragmentation events remains constant in this process. Notably, after repeated rounds of dynamic renormalisation, the kinetic equation converges to a self-similar (critical) form, where the fluctuations in cluster sizes are dominated solely by a balance between merger and fragmentation events (Supplementary Theory), while the influence of other processes becomes vanishingly small,

$$\partial_\tau f(\rho,\tau) \approx \varphi' L_{\text{fragmentation}}[f(\rho,\tau)] + \mu' L_{\text{merging}}[f(\rho,\tau)],$$

where $\varphi'$ and $\mu'$ are rescaled parameters and $\tau$ is a rescaled time (Supplemental Theory). Intuitively, this means that, as the organ grows, different sources of variance contribute to the cluster size distribution by different degree (Fig. 2B and S1A). Crucially, in the long term, contributions relating to cell fate behaviour (e.g. cell division or loss) become dominated by merger and fragmentation processes, resulting in information on the former becoming erased (Supplementary Theory). Therefore, while cell fate decisions affect the mean cluster size, the shape of the distribution is determined entirely by merger and fragmentation events (Fig. 2C), leading to the emergence of scaling behaviour observed in heart development (Fig. 1F).

Importantly, these results suggest not only that the cluster size distribution is entirely determined by its average (scaling), but also that the shape of the distribution is independent of the biological context (universality). The form of the scaling function, $\phi$, relies on the dependence of the merging and fragmentation rates on cluster size. In a uniformly growing tissue, clone merger and fragmentation events are the result of the slow diffusive motion of clusters originating from random forces exerted by the surrounding tissue (*17*). In this case, the resulting scaling form is well-approximated by a log-normal size dependence (Fig. 2C, Supplementary Theory). Indeed, such distributions are typical of merging and fragmentation processes and describe the empirical distribution of droplet sizes in aerosols (*18*, *19*). Similar universal behaviour is recapitulated by a simple lattice-based Monte Carlo simulation of uniform tissue growth, where the stochastic nature of cell division alone leads to merger and fragmentation (Figure S1B and Supplementary Theory). Importantly, this analysis provides an explanation for the observed scaling behaviour of labelled cluster sizes of mouse heart, where the distribution indeed follows a strikingly log-normal size dependence (Fig. 3A,B and S2A,B). To further challenge the universality of the scaling dependences, we used a similar genetic labelling strategy to trace the fate of early developmental precursors in mouse liver and pancreas as well as the late stage development of zebrafish heart (*20*). In all cases, cluster size distributions showed collapse onto a log-normal size dependence (Fig. 3C-F and S2C-E), with the notable exception of a subpopulation of pancreatic precursors (see below).

This analysis shows that, in the long term, the collective cellular dynamics leads to a critical state dominated by a balance between merging and fragmentation events. The emerging universal scaling distributions progressively become void of information on underlying biological processes on a time scale determined by the merging and fragmentation rates. But how can such information be recovered? In analogy to the turnover of adult homeostatic tissues, such as interfollicular epidermis or intestine (*4*, *21*), the behaviour of the size distribution under renormalization (Fig. 2B and Supplemental Theory) shows how lineage-specific information can be recovered: First, it is preserved in the non-universal cluster size dependences at short times post-labelling, prior to convergence to the scaling regime. Second, convergence onto universal scaling dependences is the slowest for small cluster sizes ($x \ll \langle x \rangle$). Third, if the rate of clone merger is negligibly small, different cluster size distributions can emerge according to the mode of cell division. The range of possible behaviours is summarised in Table 1. Finally, as merging and fragmentation are emergent properties of cell fate decisions, deviations from the scaling form can inform on structural properties of organ formation. As an example, in the developing pancreas, acinar cells initiate from precursors localized at the tips of a complex ductal network and aggregate as cohesive cell clusters thereby supressing clonal fragmentation. This results in a departure from scaling behaviour of the cluster size distribution (Figs. 3F and S2F).

In recent years, there has been a growing emphasis on genetic lineage tracing as a tool to resolve the proliferative potential and fate behaviour of stem and progenitor cells in normal and diseased tissues (*1*). Here, we have shown that the collective cellular dynamics in tissue growth and turnover lead to universal clone dynamics, where cluster size distributions become independent of the fate behaviour of cell populations. As well as highlighting the benefit of low-density labelling and the dangers of making an unguarded assessment of clonality in lineage labelled systems, these findings identify quantitative strategies to unveil cell fate-specific information from short-term or small cluster size dependencies, with potential applications to studies of clonal dynamics in both healthy and diseased states. At the same time, by highlighting the unexpected emergence of core concepts of statistical physics in a novel context, this study provides a model of how the cellular dynamics of living tissues can serve as a laboratory for statistical physics.

## Methods

### Surface area analysis of mosaically labeled hearts

To generate mosaically labelled hearts at high density, *Mesp1-Cre* mice (*22*) were crossed with the Rosa-Confetti reporter mice (*23*) kindly provided by Hans Clevers. Hearts collected at embryonic days E12.5 and P1 were fixed in 4% paraformaldehyde for 1hr at room temperature. Nuclei were counterstained with Topro3 (1/500, Invitrogen). The surface images were acquired with a confocal microscope (LSM780; Carl Zeiss). The surface area (SA) of each independent clusters was measured using Fiji software (*24*) on the maximum intensity projection.

### Pancreas

R26R-CreERT2; R26-Confetti mice were intraperitoneally injected with Tamoxifen (from Sigma) at 0.030mg per gram of female at E12.5 of pregnancy under Home Office guidelines, Animal Scientific Procedure Act (ASPA) 1986. P14 pancreas was fixed in 4% Paraformaldehyde (PFA) overnight, and then washed in PBS. Samples were sucrose-treated (30%) and mounted in OCT, and subsequently thick 100µm cryostat sectioned. Sections were rehydrated in PBS, blocked overnight in PBS, 2% donkey serum and 0.5% Triton-100X. The samples were incubated in Dolichos biflorus agglutinin (DBA), biotinylated (from Vectorlabs) for 3 days at 4°C, and AF647-Streptavidin (from Life Technologies) was applied for 2 days at 4°C. Next, sections were cleared with RapiClear 1.52 (from SunJin Lab). Images were acquired with Leica TCS SP5 confocal microscope, using the tiling mode. The images were analyzed with Volocity and volumes and coordinates of centers of clonal clusters quantified. To obtain 3D reconstructions from Z stacks obtained with Leica SP5 microscope, Imaris (v8, Bitplane) was used.

**Liver**

R26R-CreERT2$^{+}$;Rainbow$^{+}$ mice were a kind gift from Magdalena Zernicka-Goetz (University of Cambridge, UK). R26R-CreERT2$^{+}$;Rainbow$^{+}$ male mice were crossed with wild-type MF1 females and labelling induced by intraperitoneal injection of pregnant dams with Tamoxifen (Sigma). Tamoxifen was prepared at 10 mg/mL in sunflower oil and induction performed using 0.025 mg Tamoxifen per gram of pregnant dam. Pregnant dams were induced at E9.5 and the resulting pups had livers collected at postnatal day P30 – P45. Livers were divided into pieces of thickness ~10mm, washed at least 3 times in PBS to remove blood and fixed in 4% Paraformaldehyde overnight before being washed twice in PBS. Liver pieces were mounted in 4% Low Melt Agarose (Bio-Rad) and 100µm thick sections cut using a vibratome (Leica VT1000 S). Thick sections were stored in PBS at 4 °C before immunostaining. Briefly, sections were blocked in PBS + 5% DMSO (Sigma) + 2% donkey serum (Sigma) + 1% Triton-X100 (Sigma) overnight before incubation in PBS + 1% DMSO + 2% donkey serum + 0.5% Triton-X100 + 1:40 goat anti-Osteopontin (R&D Systems, AF808) for 3 days at 4 °C. Following several washes in PBS + 1% DMSO + 0.5% Triton-X100 at 4 °C for 24 h, sections were incubated in PBS + 1% DMSO + 2% donkey serum +

0.5% Triton-X100 + 1:250 donkey anti-goat antibody conjugated to AF647 (Life Technologies) for 2 days at 4 °C. Following the staining, sections were cleared by increasing glycerol gradient before incubation with PBS + 1:1000 Hoechst 33342 (Sigma) for 1h at 4 °C to counterstain nuclei and mounted with Vectashield (Vector Laboratories). Images of liver sections were acquired using a Leica TCS SP5 confocal microscope and processed using LAS AF Lite software (Leica). Cell numbers for each labelled cluster were counted manually from acquired images.

**Code availability**

Custom code used to in this study is available from the corresponding authors upon reasonable request.

## Data availability

The data that support the plots within this paper and other finding of this study are available from the corresponding author upon request.

**Acknowledgements:**

B.D.S. acknowledges the support of the Wellcome Trust (grant number 098357/Z/12/Z). F.L. is supported by a long-term EMBO fellowship and the postdoctoral fellowship of the FNRS. S.C. is supported by a FRIA/FNRS fellowship. C.B. is supported by the ULB, a research grant of the FNRS, the foundation Bettencourt Schueller, the Foundation ULB and the Foundation Baillet Latour. M.S. is supported by an MRC doctoral training award and A.P. is supported by MRC Research grant MR/K018329/1. M.H. is a Wellcome Trust Sir Henry Dale Fellow and is jointly funded by the Wellcome Trust and the Royal Society (104151/Z/14/Z); M.H.and N.P. are funded by a Horizon 2020 grant (LSFM4LIFE). C.H. was funded by a Cambridge Stem Cell Institute Seed funding award for interdisciplinary research awarded to M.H. and B.D.S.. We are grateful to Kenneth D. Poss and Vikas Gupta for making a digital version of their data available to us.

**Author contributions:**

S.R. and B.D.S. conceived the project. S.C., F.L., C.J.H., N.P., and M.S. performed the experiments and collected the raw data. M.H. supervised the liver experiments. S.R. developed the theory, performed the modelling and statistical analysis. S.R. and B.D.S drafted the manuscript. All authors edited and approved the final manuscript.

**Competing financial interests:** The authors declare no competing financial interests.

**Data availability statement:** The data that support the findings of this study are available from the corresponding author upon reasonable request.

**Ethical statement:** We have complied with all relevant ethical regulations. Mesp1-Cre mice colonies were maintained in a certified animal facility in accordance with European guidelines. These experiments were approved by the local ethical committee under the reference #LA1230332(CEBEA). Research using mice for pancreas and liver samples has been regulated under the Animal (Scientific Procedures) Act 1986 Amendment Regulations 2012 following ethical review by the University of Cambridge Animal Welfare and Ethical Review Body (AWERB). These experimental data sets were obtained as by-products from other research projects undertaken by the respective laboratories.

Fig. 1

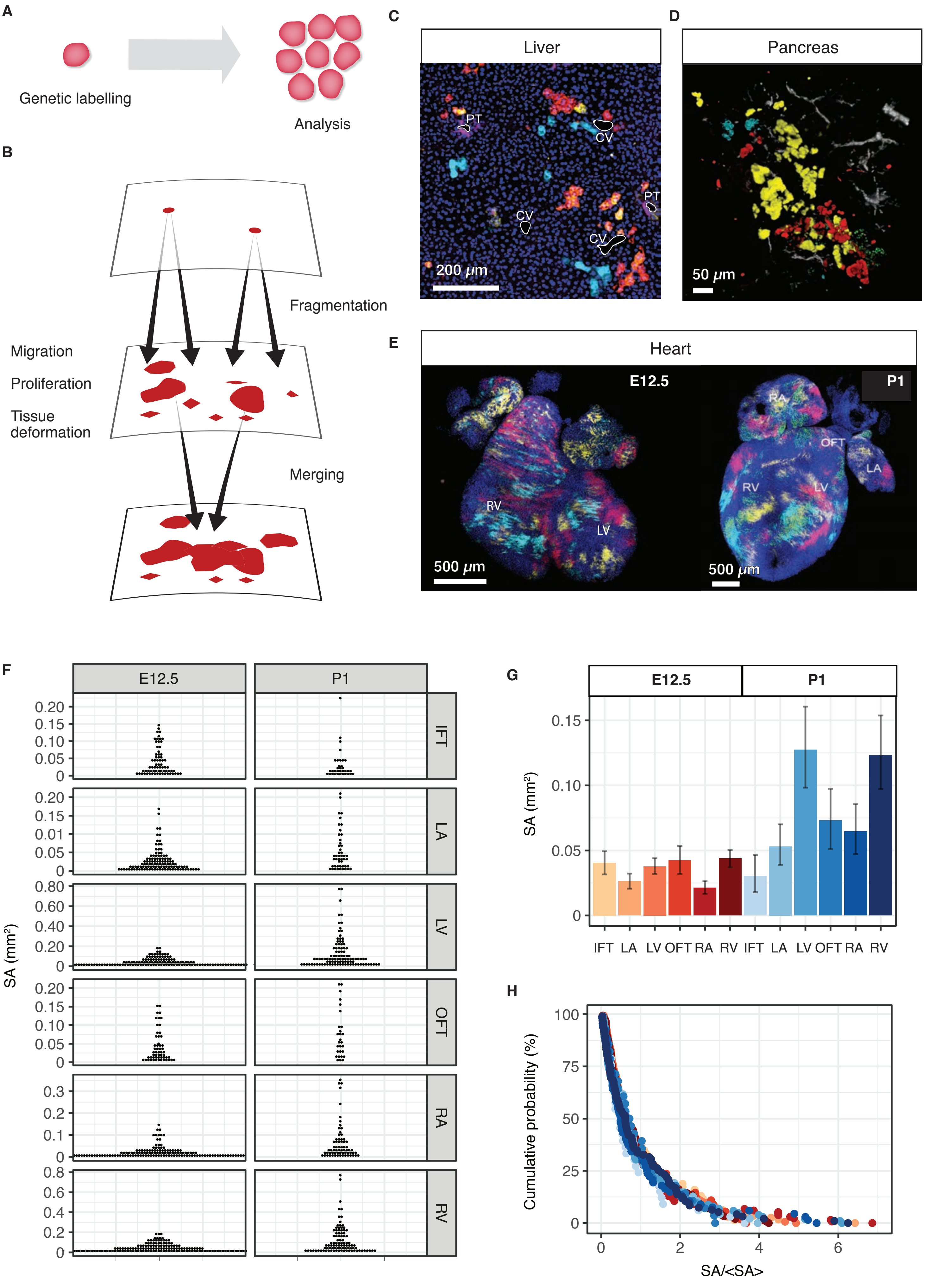
A
Genetic labelling
Analysis
B
Fragmentation
Migration
Proliferation
Tissue deformation
Merging
C
Liver
PT
CV
200 μm
D
Pancreas
50 μm
E
Heart
E12.5
P1
RA
OFT
LA
RV
LV
500 μm
F
E12.5
P1
IFT
LA
LV
OFT
RA
RV
SA (mm²)
G
SA (mm²)
IFT LA LV OFT RA RV IFT LA LV OFT RA RV
H
Cumulative probability (%)
SA/<SA>

**Fig. 1. Clonal dynamics during tissue development. (A)** Lineage tracing allows resolving clonal dynamics using a "two-time" measurement in living organisms. **(B)** Merger and fragmentation of labelled cell clusters occur naturally because of large-scale tissue rearrangements during the growth and development of tissues. **(C,D)** Illustration of clone fragmentation in mouse during the development of **(C)** liver and **(D)** pancreas (collection at post-natal day (P)45 and P14, respectively) following pulse-labelling using, respectively, R26R-CreERT2;Rainbow and R26R-CreERT2; R26-Confetti at E9.5 and E12.5, respectively. Portal tracts (PT) and central veins (CV) are highlighted in white, osteopontin (a ductal marker) is shown in purple and nuclei are marked in blue. Pancreatic ducts are shown in grey. **(E)** High density (mosaic) labelling of mouse heart using the Mesp1-Confetti system showing the left/right atrium (L/RA), left/right ventricle (L/RV) and the in/out-flow tracts (I/OFT). **(F)** Distributions of cell cluster sizes on the surface of the developing mouse heart at E12.5 (680 clusters from 4 mice) and P1 (373 clusters from 3 mice). **(G)** Average cluster sizes in different heart compartments and time points during development. Error bars denote 95% confidence intervals. **(H)** Rescaled cluster size distributions showing scaling behaviour.

**Fig. 2**

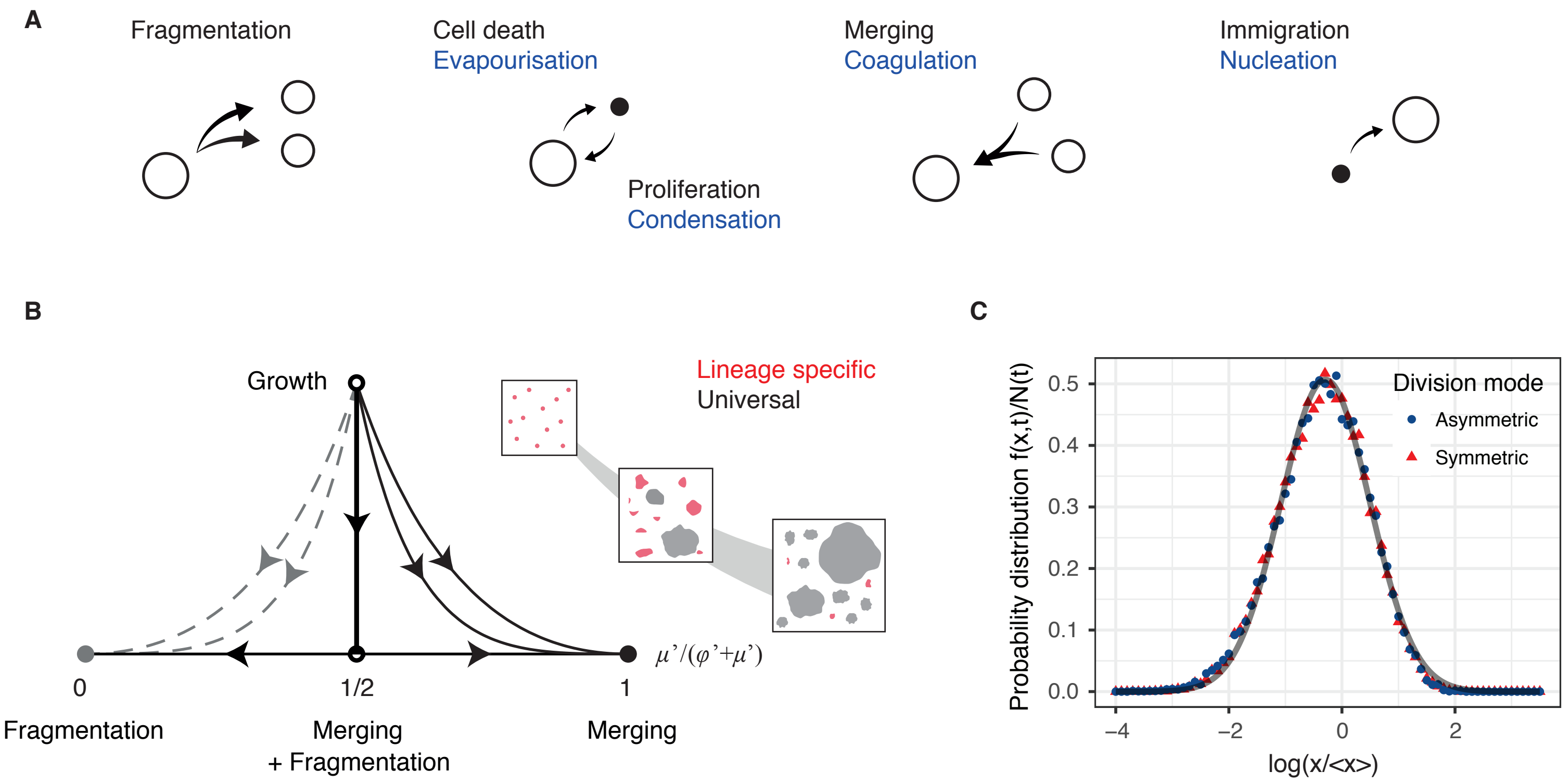

**Fig. 2. Origin of scaling and universality in clonal dynamics during development. (A)** Sizes of labelled cell clusters in developing tissues are determined by processes analogous to the kinetics of droplets in aerosols, as depicted. **(B)** Sketch of the renormalisation flow diagram showing how the relative contributions of different processes to the cluster size distribution evolve during development. At long times and/or larger cluster sizes, the time evolution of the cluster size distribution becomes controlled by three fixed points (dependent on the details of the merging and fragmentation processes), where it acquires a universal scaling dependence (Supplementary Information). The inset shows a schematic of the renormalization process, with the largest cluster sizes (grey) converging more rapidly onto the universal distribution than the smallest cluster sizes (red). **(C)** Rescaled cluster size distributions for different division modes obtained by numerical simulations (Supplemental Theory) collapse onto a universal log-normal form (grey line).

Fig. 3

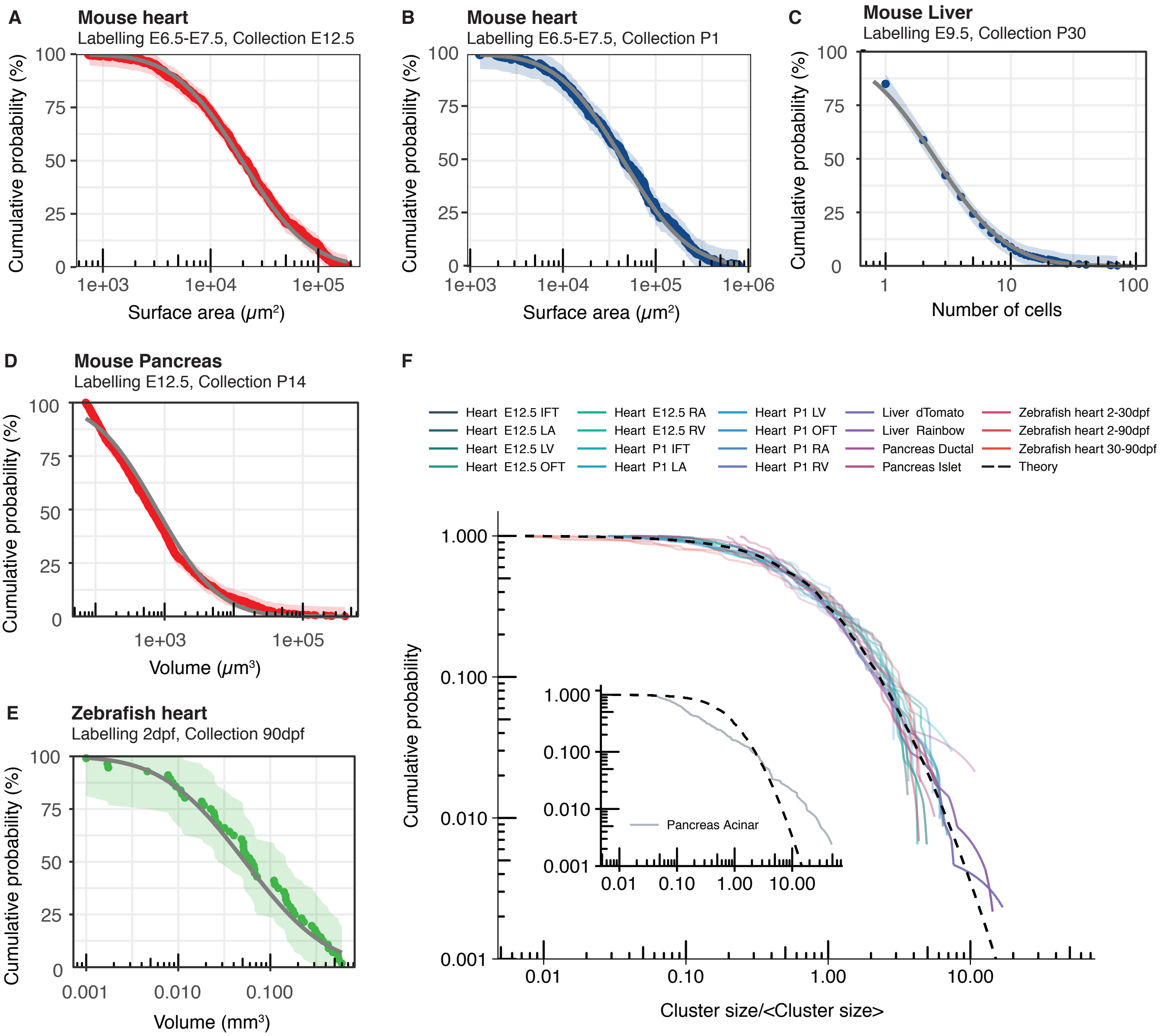

**Fig. 3. Universality of cluster sizes in different tissue types and organisms. (A-B)** Cumulative cluster size distributions obtained from lineage tracing studies of the mouse heart. **(C-E)** Experimental cumulative cluster size distributions for **(C)** mouse liver (892 clusters from 4 mice), **(D)** mouse pancreas (988 clusters from 3 mice), and **(E)** zebrafish heart (from (*20*)) collapse onto the predicted universal log-normal dependence fitted by maximum likelihood estimation (grey). Data shown in colour and shading shows 95% Kolmogorov confidence intervals. **(F)** Experimental cumulative cluster size distributions (solid lines) separated by time, region, cell type labelling strategy collapse onto a universal shape (dashed line) with the exception of a subset of pancreatic acinar cells (inlay).

**Table 1.**

| **Growth mode** | **Clonal** | **Fragmentation** | **Merging & fragmentation** |
|---|---|---|---|
| Exponential | $\langle x\rangle^{-1}\exp\left(-x/\langle x\rangle\right)$<br><br>$\langle x\rangle = \exp(t)$ | $\varphi\exp\left(-\varphi^{-1}x\right)$<br><br>$\langle x\rangle = \varphi^{-1}$ | $\begin{cases} (x/\langle x\rangle)^{\alpha} & x \ll \langle x\rangle \\ \exp(-x/\langle x\rangle) & x \gg \langle x\rangle \end{cases}$<br><br>$\langle x\rangle = \exp(t)$ |
| Linear | $\frac{1}{\sqrt{2\pi\langle x\rangle}}\exp\left[-\frac{(x-\langle x\rangle)^2}{2\langle x\rangle}\right]$<br><br>$\langle x\rangle = t$ | $\varphi x\left(2+\sqrt{\varphi}x\right)\exp\left(-\sqrt{\varphi}x-\frac{\varphi}{2}x^2\right)$<br><br>$\langle x\rangle = \varphi^{-1/2}$ | $\begin{cases} (x/\langle x\rangle)^{\alpha} & x \ll \langle x\rangle \\ \exp(-x/\langle x\rangle) & x \gg \langle x\rangle \end{cases}$<br><br>$\langle x\rangle = t$ |
| Homeostasis | $\langle x\rangle^{-1}\exp\left[-x/\langle x\rangle\right]$ (see Ref. 6)<br><br>$\langle x\rangle \propto t$ | $J(x)$ (see Ref. 24)<br><br>$\langle x\rangle = \text{const.}$ | $\begin{cases} (x/\langle x\rangle)^{\alpha} & x \ll \langle x\rangle \\ \exp(-x/\langle x\rangle) & x \gg \langle x\rangle \end{cases}$<br><br>$\langle x\rangle = \text{const.}$ |

**Table 1. Non-universal dependencies of the cluster size distribution.** Analytical expressions for the cluster size distribution (top row in each cell) and average cluster size (bottom row). Shown are expressions in situations, where labelling density is clonal, labelling density is almost clonal but clones are subject to fragmentation, and where both merging and fragmentation of clones occur (left to right). As merging and fragmentation both result from tissue rearrangements merging should always imply fragmentation. Time is measured in units of the cell cycle time. Expressions are valid after convergence to the scaling regime, when the typical cluster size is much larger than the size of single cells, and in the mean-field limit, which is a good approximation for two and three dimensional tissues. In addition, it is assumed that the full spectrum of cluster sizes can be experimentally resolved. If clones fragment but not merge fragmentation and growth ultimately compensate to lead to a stationary distribution. In case of clonal merging and fragmentation expressions give empirical approximations, where $\alpha$ depends on the details of the merging and fragmentation processes (see Supplemental Theory).

**Fig. S1**

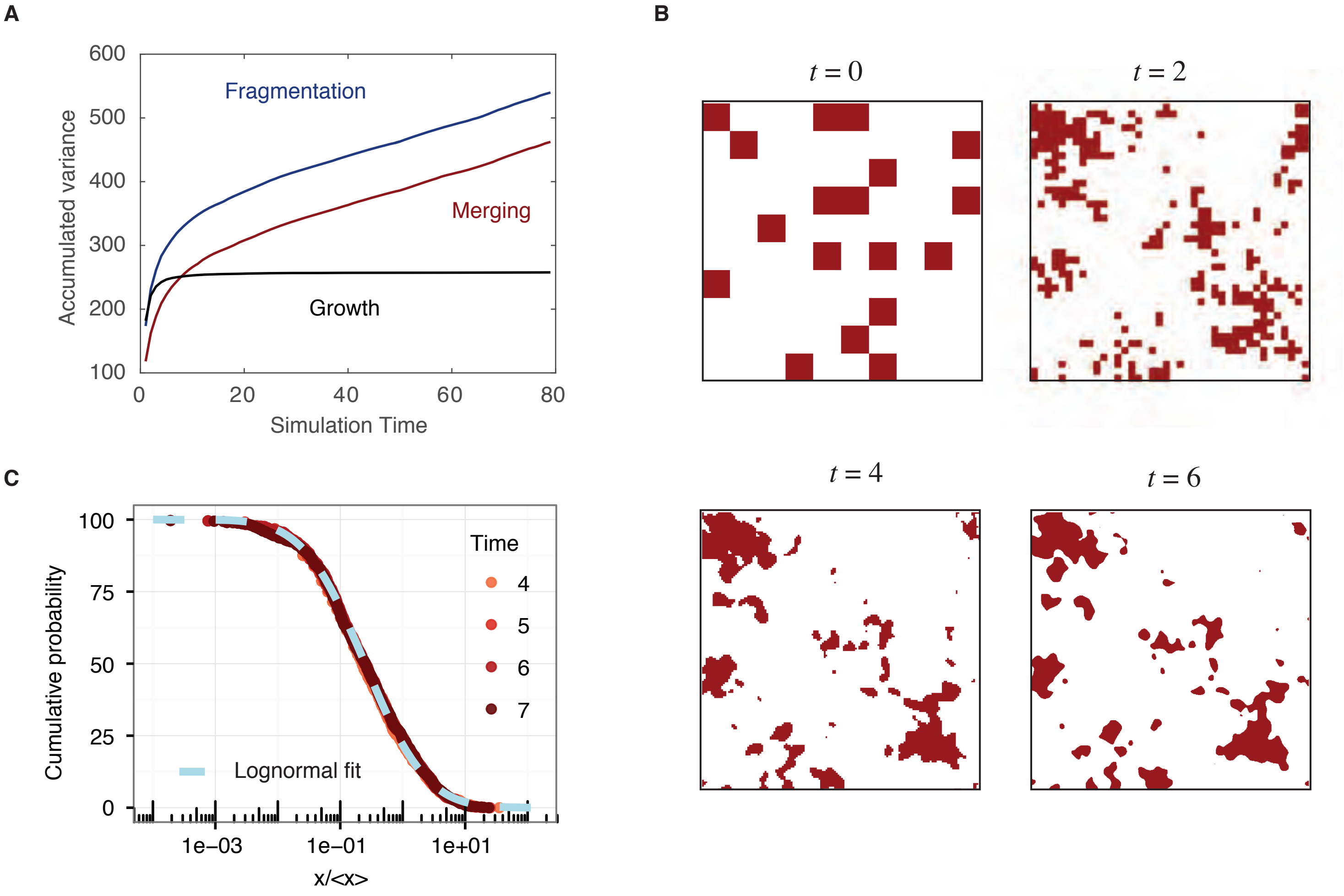

**Fig. S1. Numerical simulations demonstrate the emergence of universal scaling behavior.** **(A)** Accumulated contributions to the variance of the cluster size distributions stemming from different processes. Over time variance is dominated by merging and fragmentation processes. **(B)** Snapshots of numerical simulations of marked clones embedded into a growing cell population (slice through a cubic lattice). **(C)** Rescaled cumulative distributions of cluster sizes obtained from the lattice simulations. For details of the numerical implementations and parameter values see Supplemental Theory.

Fig. S2

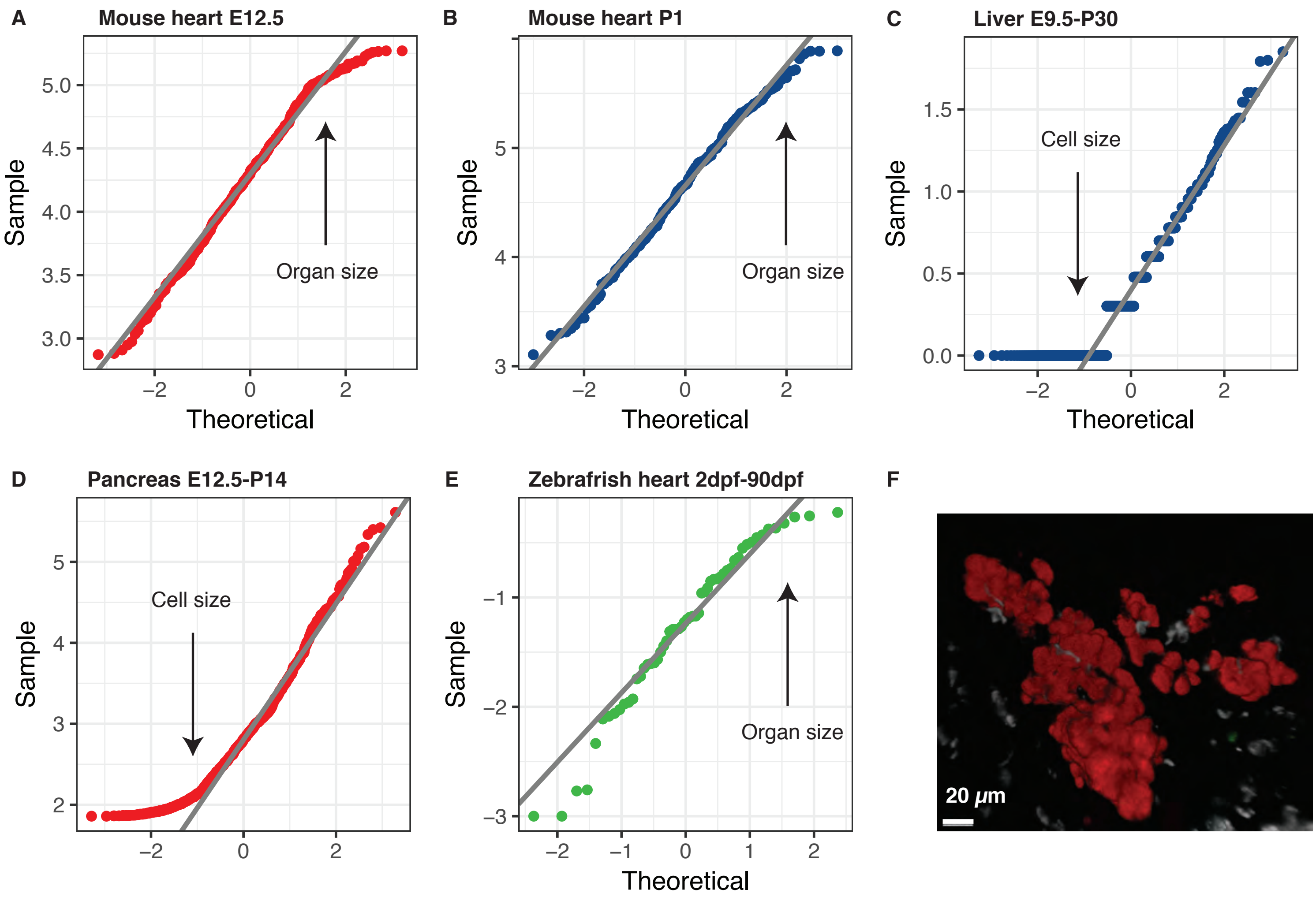

**Fig. S2. Comparison of the empirical cluster size distribution in various tissues using quantile-quantile plots.** Quantile-quantile plots for **(A,B)** mouse heart, **(C)** mouse liver, **(D)** pancreas and **(E)** zebrafish heart. Plotted are the theoretical quantiles versus the empirical quantiles of the log-transformed data. Log-normality is indicated by a straight, diagonal line. Necessary deviations from log-normality occur where cluster sizes are of the order of the size of single cells or the whole organ (indicated by arrows). **(F)** Example of a cohesive cluster of acinar cells (red) localised at the tips of a ductal network (white). Cells were labelled using R26R-CreERT2; R26-Confetti at E12.5 and collected at P14.

# SUPPLEMENTAL THEORY

STEFFEN RULANDS & BENJAMIN D. SIMONS

In this Supplemental Theory we present details of the calculations used to infer the results presented in the main text. This document is structured as follows: By renormalising the mean-field dynamics, we first show that the kinetics of clone merger and fragmentation give rise asymptotically to scaling behaviour and universality in the size distributions of labelled cell clusters. By deriving specific functional forms of the merger and fragmentation kernels, we then infer the shape of the scaling distribution. Guided by these results, we then identify strategies to resolve non-universal, lineage specific, dependencies and give analytical expressions for non-universal size distributions in two generic examples.

# CONTENTS

# 1 EMERGENCE OF SCALING BEHAVIOUR

Transgenic mouse models allow the inducible hereditary labelling of targeted subsets of cells in tissues by activating fluorescent reporter genes. The expression of this label is inherited by all progeny of a marked cell (which together constitute a clone). While static measures based on the size and composition rarely allow fate behavior to be inferred from an individual clone, information can often be recovered from the properties of a statistical ensemble of clones. While, under homeostatic conditions, the progeny of a marked cell typically form spatially cohesive groups, in conditions of development or cancerous growth, large-scale tissue deformations and cell migration can lead to the fragmentation of clones leading to ambiguities in clonal assignments. Similarly, initially disconnected distinct clones may merge to form larger compounds of commonly labelled cells. Under conditions of merger and fragmentation, can lineage specific information be recovered from such clonal data? The empirical observation of robust scaling behaviour in clonally labelled heart, where a single scale defines the distributions of the sizes of such labelled clusters, suggests a simple and universal underlying mechanism.

To understand the emergence of scaling behaviour of the labeled cell cluster distribution and to infer its shape, we begin by defining the number $f(x,t)$ of clusters with a given size, $x$, at time, $t$, post-labelling. More precisely, $f(x,t)\mathrm{d}x$ defines the average number of cell clusters whose "masses" lie between $x$ and $x+\mathrm{d}x$. Depending on the experimental context, the mass or size, $x$, might refer to the sectional or surface area covered by a cluster of labelled cells, the number of cells in a cluster or, in the case of full three-dimensional reconstructions of tissue sections, the volume of the cluster. At this "mesoscopic" level of description, the time evolution of the distribution $f(x,t)$ is governed by multiple processes, including cell division, cell death, differentiation, clone merger and clone fragmentation. For simplicity, here we focus on growth, merger and fragmentation. Importantly, the analysis of other processes is analogous, as emphasized briefly in the Appendix. Formally, then, the time evolution of $f(x,t)$ can be written as a sum of operators of the form

$$\partial_t f(x,t) = \mathcal{L}_{\text{growth}}[f(x,t)] + \varphi \mathcal{L}_{\text{fragmentation}}[f(x,t)] + \mu \mathcal{L}_{\text{merging}}[f(x,t)]\,. \tag{1}$$

Here, time is measured in units of the average cell division time and $\varphi$ and $\mu$ are the fragmentation and merging rates, respectively. In the continuum limit, the growth term is of the form [1]

$$\mathcal{L}_{\text{growth}}[f(x,t)] = -\frac{\partial}{\partial x}[x^\beta f(x,t)]\,. \tag{2}$$

The exponent $\beta$ characterizes the mode of cell divisions: For example, $\beta = 1$ corresponds to a process of symmetrical self-renewal where, after division, both daughter cells remain in cell cycle and the rate of growth is proportional to the size of the cluster. By contrast, in a population of asymmetrically dividing cells, progenitor cell division gives rise to one cycling progenitor and one cell that exits cell cycle. In this case the pool of dividing cells remains constant translating to the exponent $\beta = 0$.

A change in cluster frequency of a given size, $x$, due to fragmentation may occur either through fragmentation of clusters of size $x$ into smaller sizes or the fragmentation of larger clusters into subclusters of size $x$. We denote by $F(x', x'')$ the rate of fragmentation events, where a cluster of size $x' + x''$ gives rise to fragments of sizes $x'$ and $x''$. With the *fragmentation kernel* so defined, the fragmentation operator applied to $f(x,t)$ takes the form [2]

$$\mathcal{L}_{\text{fragmentation}}[f(x,t)] = 2\int_0^\infty F(x,x')f(x+x',t)\mathrm{d}x' - f(x,t)\int_0^x F(x-x',x')\mathrm{d}x' \,.$$

The factor of 2 reflects the fact that each fragmentation event gives rise to two clusters.

Similarly, the frequency of clusters of size $x$ may decrease through the merger of clusters of size $x$ with a cluster of equivalent colour and size $x'$ to produce a cluster of size $x + x'$, or it may increase through the merger of a cluster of $x'$ with a cluster of size $x - x'$. Again, the rate of mergers of clusters of sizes $x$ and $x'$ is defined by a *merging kernel*, $K(x, x')$, such that the average number of mergers between clusters of size $x$ to $x + \mathrm{d}x$ and those of size $x'$ to $x' + \mathrm{d}x'$ is $K(x,x')f(x,t)f(x',t)\mathrm{d}x\mathrm{d}x'\mathrm{d}t$ during the time interval $t$ to $t + dt$. With this definition, the terms arising from mergers take the form [1]

$$\begin{aligned}\mathcal{L}_{\text{merging}}[f(x,t)] = &\int_0^x K(x',x-x')f(x,t)f(x-x',t)\mathrm{d}x' \\ &- f(x,t)\int_0^\infty K(x,x')f(x',t)\mathrm{d}x' \,.\end{aligned} \tag{3}$$

In summary, the distribution of clusters of a given size evolves according to a *mean-field Master equation* of the form

$$
\begin{aligned}
\frac{\partial}{\partial_t} f(x,t) = &-\frac{\partial}{\partial x}[x^\beta f(x,t)] \\
&+ \varphi \left[ 2\int_0^\infty F(x,x') f(x+x',t) \mathrm{d}x' - f(x,t) \int_0^x F(x-x',x') \mathrm{d}x' \right] \\
&+ \mu \left[ \int_0^r K(x',x-x') f(x,t) f(x-x',t) \mathrm{d}x' \right. \\
&\qquad \left. - f(x,t) \int_0^\infty K(x,x') f(x',t) \mathrm{d}x' \right] \\
&+ \text{ additional terms.}
\end{aligned} \tag{4}
$$

Importantly, we will argue that the additional processes provide subleading contributions to the distribution, which can be formally neglected in the scaling limit. Using similar arguments, the size distribution, $g(y,t)$, of tissue not labelled in the given colour evolves as

$$
\frac{\partial}{\partial_t} g(y,t) = -\frac{\partial}{\partial y} \left[ y^\beta g(y,t) \right]. \tag{5}
$$

By describing the time evolution of labelled cluster sizes in such a manner we made two assumptions. First, we defined labelled clusters entirely by their size; the shape of clusters must either be neglected or taken into account by making an appropriate choice for the merger and fragmentation kernels. Second, we assumed that spatial correlations are negligible and employed a mean-field approximation. This is justified by the fact that the critical dimension of merger-fragmentation type processes is below one [3].

### 1.1 Renormalisation of the kinetic equations

The solution of equations like (4) is rarely feasible, even for trivial choices of the kernels $F(x, x')$ and $K(x, x')$. However, in common with droplets in aerosols, lineage tracing assays, particularly in a developmental context, give rise to clusters of labelled cells in tissues that typically span multiple orders of magnitude in size. Due to technical limitations in microscopy, it is rare that cluster sizes can be quantified simultaneously and equally well across all of these length scales. Instead, the measured size distribution of labelled clusters is usually dominated by statistical fluctuations due to cluster dynamics on large scales, while small scale events may not be resolved. To make analytical progress in understanding the emergence of scaling of the cluster size distribution, it is therefore sufficient to study the impact of large-scale fluctuations.

In statistical physics, large-scale fluctuations are typically studied by successively coarse-graining some degrees of freedom in a given system and then monitoring how this procedure affects fluctuations from different origins – a theoretical strategy known as *renormalisation group*. Here, we follow a conceptually similar approach to understand the origin of scaling behaviour in lineage tracing assays. To this end, we employ a dynamic renormalisation strategy: To identify the kinetic processes dominating large-scale fluctuations, we repeatedly coarse-grain cluster sizes and developmental times. We will see that, under this procedure, the dy-namics asymptotically converges to a "critical" process, which is dominated by the kinetics of merger and fragmentation.

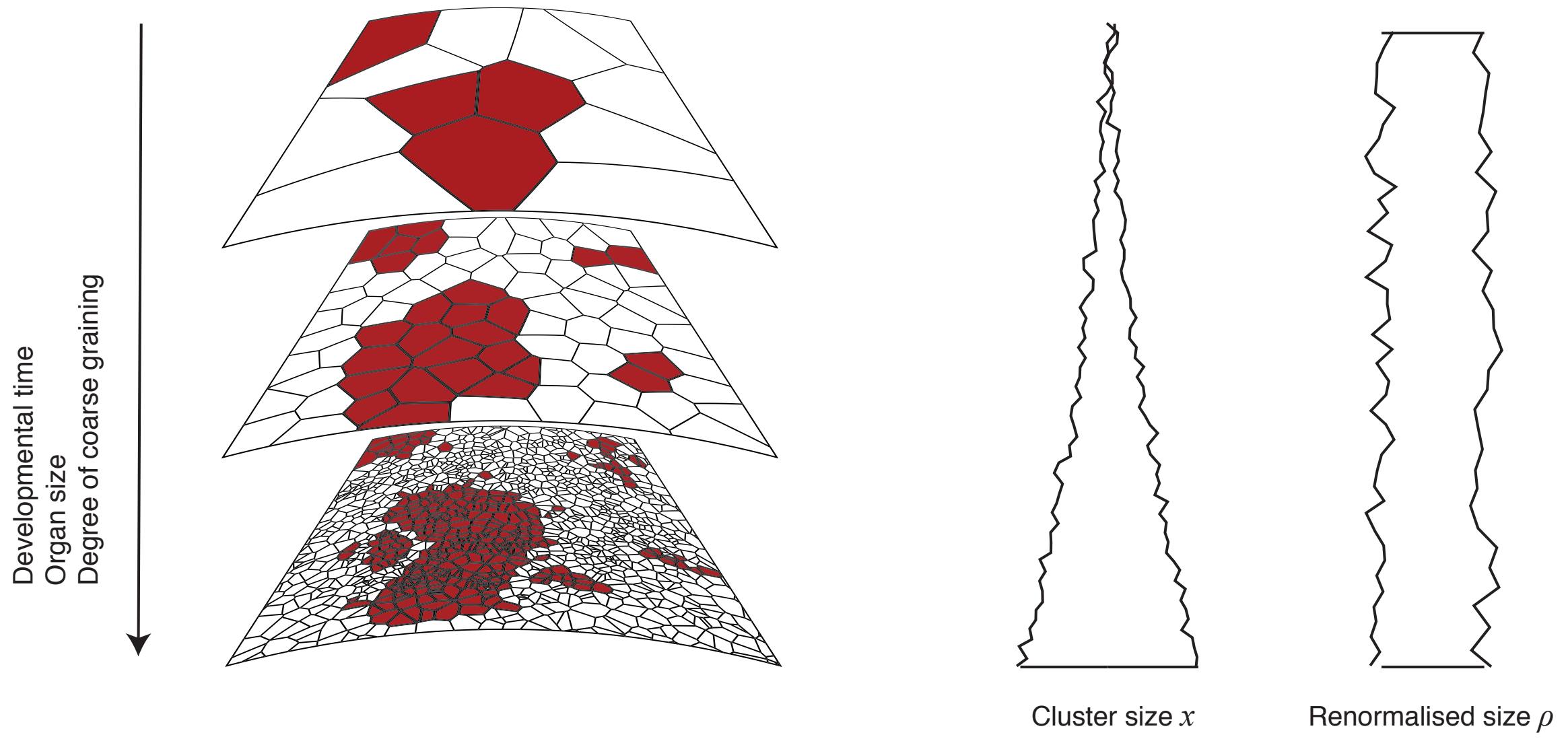

**Figure 1:** Illustration of the dynamic coarse graining procedure. The rescaling compensates the overall growth of the tissue (left). While in the presence of merging and fragmentation typical cluster sizes increase in size over time, the characteristic renormalised size is time independent (right, illustration of a typical stochastic realisation).

Specifically, we choose a renormalisation scheme that eliminates the time derivative of the first moment of the cluster size distribution (Figure 1). We begin by considering the sizes of two subpopulations of cells: The size of labelled clusters, $x$, and the size of the remaining part of the growing tissue, $y$. Whenever a cell in the tissue divides we renormalise all cluster sizes by the amount of increase in the tissue size:

$$x \to \frac{x}{1 + \mathrm{d}X} \equiv \rho \cdot \langle x \rangle (t = 0)\,, \tag{6}$$

with $\mathrm{d}X$ denoting the differential increase in tissue size, $X \equiv x + y$, and $\langle x \rangle (t = 0)$ being the average initial size of labelled clusters. This rescaling implicitly defines the relative rescaled cluster sizes, $\rho$. In the following we will express the kinetics in terms of these renormalised coordinates, $\rho$. To begin we first note that $\rho$ is the fraction of a tissue that is occupied by a given labelled cluster in a specific realisation of the stochastic process. In other words, we can write $\rho = x/X$.

To understand how the kinetic equations (4) behave under renormalization, we first focus on the growth dynamics alone. Since the mean-field Master equation is defined phenomenologically, we can derive the time evolution equation in the renormalised coordinates by phenomenological arguments as well. Measuring sizes in units of the size of a single cell, the time evolution of the joint distribution of $\rho$ and $X$ is governed by two processes:

1. A cell which is not part of a given cluster divides such that the relative fraction of the labeled cluster decreases multiplicatively, $\rho \to x/(X+1) = \rho/(1+1/X)$. This process occurs at a rate that scales in proportion to the number of cells outside the cluster, $[X(1-\rho)]^{\beta}$.
2. A cell within a given cluster divides, yielding a multiplicative contribution from the expansion of the tissue and an additive contribution from the growth of the cluster, $\rho \to (x+1)/(X+1) = \rho/(1+1/X) + 1/(1+X)$. In expand-ing tissues, the rate of cell divisions is proportional to the size of the cluster, $(X\rho)^\beta$.

With these definitions, the time evolution of the number of clusters of relative size $\rho$ in organs of size $X$, $\psi(\rho, X, t)$, follows a Master equation of the form

$$\begin{aligned}\frac{\partial}{\partial t}\psi(\rho, X, t) = X^\beta\left(\rho - \frac{1}{X}\right)^\beta \psi\left[\left(\rho - \frac{1}{X}\right)\left(1 + \frac{1}{X-1}\right), X-1, t\right] \\ - (\rho X)^\beta \psi(\rho, X, t) \\ + X^\beta\left(1 - \rho - \frac{1}{X}\right)^\beta \psi\left[\rho\left(1 + \frac{1}{X-1}\right), X-1, t\right] \\ - X^\beta(1-\rho)^\beta\psi(\rho, X, t)\,.\end{aligned} \tag{7}$$

By definition of the rescaling, it is clear that the first moment associated with the growth process is constant, $\frac{\partial}{\partial t}\int_0^1 \rho\psi(\rho, X, t)\mathrm{d}\rho = 0$. To the next highest order, the growth dynamics in the rescaled coordinates can therefore be approximated by a diffusive process. To formalise this and to estimate the contribution of the growth dynamics to higher moments, we formally expand in the step sizes of the stochastic process,

$$\frac{\partial}{\partial t}\psi(\rho, X, t) = \sum_{n=1}^{\infty}\frac{(-1)^n}{n!}\frac{\partial^n}{\partial_\rho^{\alpha_\rho}\partial_X^{\alpha_X}}\left[R^{\alpha_\rho,\alpha_X}(\rho, X)\psi(\rho, X, t)\right], \tag{8}$$

where the jump moments are defined as

$$R^{\alpha_\rho,\alpha_X}(\rho, X) = \int_{-\infty}^{\infty}\mathrm{d}\delta_\rho\,\mathrm{d}\delta_X\,\delta_\rho^{\alpha_\rho}\delta_X^{\alpha_X}\mathcal{W}(\rho, X; \delta_\rho, \delta_X)\,. \tag{9}$$

Here, $\mathcal{W}(\rho, X; \delta_\rho, \delta_X)$ is the transition rate for jumps of size $\delta_\rho$ and $\delta_X$ and, from Equation (7), it follows that it is of order $X^\beta$. The jump sizes scale as $\delta_\rho \in \mathcal{O}[X^{-1}]$ and $\delta_X \in \mathcal{O}[1]$, such that $R^{\alpha_\rho,\alpha_X}(\rho, X) \propto X^{\beta-\alpha_\rho}$. From symmetry, and by the defini-tion of $\rho$, it is clear that jump moments with uneven powers in $\alpha_\rho$ vanish, such that the lowest order contributions are given by

$$\begin{aligned}\frac{\partial}{\partial t}\psi(\rho,X,t) = &\left[-\frac{\partial}{\partial X}R^{0,1}(\rho,X) + \frac{1}{2}\frac{\partial^2}{\partial X^2}R^{0,2}(\rho,X) - \dots\right]\psi(\rho,X,t) \\ &+ \left[\frac{1}{2}\frac{\partial^2}{\partial \rho^2}R^{2,0}(\rho,X) - \frac{1}{6}\frac{\partial^3}{\partial\rho^2\partial X}R^{2,1}(\rho,X)\right]\psi(\rho,X,t) \\ &+ \mathcal{O}[X^{\beta-4}]\,. \end{aligned} \tag{10}$$

We continue by making the ansatz $\psi(\rho,X,t) = \vartheta(\rho,t)\chi(X,t)$ and formally separating the dependences of $R^{2,0}(\rho,X)$ on $\rho$ and $X$, $R^{2,0}(\rho,X) = A(\rho)B(X)$. Following Equations (7) and (9), $A(\rho)$ scales with $X^{-2}$ and $B(X)$ with $X^{\beta}$. Then, integrating over $X$, we obtain the time evolution equation for the marginal distribution,

$$\frac{\partial}{\partial t}\vartheta(\rho,t) = \frac{1}{2}C(t)\frac{\partial^2}{\partial\rho^2}A(r)\vartheta(\rho,t) + \mathcal{O}[X^{\beta-4}]\,, \tag{11}$$

with $C(t) = \int_0^\infty B(X)\chi(X,t)\mathrm{d}X$. $C(t)$ scales with $X_0(t)^\beta$, where $X_0(t)$ is the mean organ size at time $t$. Therefore, fluctuations in $\rho$ stemming from cell divisions decrease with $X_0^{\beta-2}$ and, as a consequence, fluctuations associated with growth do not scale with the overall size of the expanding tissue if cells proliferate symmetrically ($\beta = 1$). If cells proliferate asymmetrically ($\beta = 0$) these terms scale inversely with the total number of cells in the expanding tissue.

How do the fragmentation and merging terms change under renormalisation? To proceed, we assume that the clone fragmentation and merging kernels are homogeneous functions of the cluster sizes, i.e. $F(\lambda x, \lambda x') = \lambda^\alpha F(x,x')$. Indeed, most realistic kernels are homogeneous. For example, if the size distribution of daughter fragments is uniform and the rate of fragmentation increases linearly with cluster size, the fragmentation kernel reads $F(x, x') = 1$ and consequently $\alpha = 0$. Then, as fragmentation events do not change the total mass of labelled cells, the fragmentation terms remain structurally invariant upon rescaling, but coarse-graining introduces a cut-off in the kernel restricting the smallest possible size of daughter fragments. This cut-off is proportional to $X_0(t)$, i.e. it is of order one in the renormalised coordinates. The fragmentation terms then renormalise as

$$\begin{aligned}\varphi' \tilde{\mathcal{L}}_{\text{fragmentation}} [\vartheta(\rho, t)] = \varphi' \bigg[ 2 \int_0^\infty F_c(\rho, \rho') \vartheta(\rho + \rho', t) \mathrm{d}\rho' \\ - \vartheta(\rho, t) \int_0^\rho F_c(\rho - \rho', \rho') \mathrm{d}\rho' \bigg] ,\end{aligned} \tag{12}$$

with the rescaled fragmentation rate

$$\varphi' = X_0(t)^{\alpha+1} \varphi \,. \tag{13}$$

Similarly, we assume that the merging kernel is homogeneous, $K(\lambda x, \lambda x') = \lambda^\gamma K(x, x')$. Again, this applies for most realistic kernels. As an example, the additive kernel, $K(x, x') = x + x'$, has $\gamma = 1$ and the multiplicative kernel, $K(x, x') = xx'$, has $\gamma = 2$. The merging terms in the rescaled coordinates also remain structurally invariant, and read

$$\begin{aligned}\mu' \tilde{\mathcal{L}}_{\text{merging}} = \mu' \bigg[ \int_0^\rho K(\rho', \rho - \rho') \vartheta(\rho, t) \vartheta(\rho - \rho', t) \mathrm{d}\rho' \\ - \vartheta(\rho, t) \int_0^\infty K(\rho, \rho') \vartheta(\rho', t) \mathrm{d}\rho' \bigg] ,\end{aligned} \tag{14}$$

where the rescaled merging rate is given by

$$\mu' = X_0(t)^{\gamma+1} \mu \,. \tag{15}$$

Hence, up to a non-linear rescaling of time, the merging and fragmentation processes remain structurally invariant under renormalisation.

In summary, the renormalised dynamics in the rescaled coordinates is governed by diffusion as well as merging and fragmentation processes,

$$\begin{aligned}\frac{\partial}{\partial t}\vartheta(\rho,t) \approx& \frac{1}{2}C(t)\frac{\partial^2}{\partial\rho^2}A(\rho)\vartheta(\rho,t)\\ &+ X_0(t)^{\alpha+1}\varphi\left[2\int_0^\infty F(\rho,\rho')\vartheta(\rho+\rho',t)\mathrm{d}\rho'\right.\\ &\left.\qquad - \vartheta(\rho,t)\int_0^\rho F(\rho-\rho',\rho')\mathrm{d}\rho'\right]\\ &+ X_0(t)^{\gamma+1}\mu\left[\int_0^\rho K(\rho',\rho-\rho')\vartheta(\rho,t)\vartheta(\rho-\rho',t)\mathrm{d}\rho'\right.\\ &\left.\qquad -\vartheta(\rho,t)\int_0^\infty K(\rho,\rho')\vartheta(\rho',t)\mathrm{d}\rho\right].\end{aligned} \tag{16}$$

Importantly, as will be discussed below, typical values for $\alpha$ and $\gamma$ are greater or equal to 0. Hence, as the organ grows and $\rho X_0(t) \to \infty$, these processes contribute with different weights to the fluctuations determining the shape of the asymptotic distribution of cluster sizes. Specifically, processes involving cell divisions, cell death or immigration do not contribute to the shape of the cluster size distribution in the asymptotic limit. In this limit, the mean field Master equation reduces to

$$\begin{aligned}\frac{\partial}{\partial t}\vartheta(\rho,t) \approx& X_0(t)^{\alpha+1}\varphi\left[2\int_0^\infty F_c(\rho,\rho')\vartheta(\rho+\rho',t)\mathrm{d}\rho'\right.\\ &\left.\qquad - \vartheta(\rho,t)\int_0^\rho F_c(\rho-\rho',\rho')\mathrm{d}\rho'\right]\\ &+ X_0(t)^{\gamma+1}\mu\left[\int_0^\rho K(\rho',\rho-\rho')\vartheta(\rho,t)\vartheta(\rho-\rho',t)\mathrm{d}\rho'\right.\\ &\left.\qquad -\vartheta(\rho,t)\int_0^\infty K(\rho,\rho')\vartheta(\rho',t)\mathrm{d}\rho\right].\end{aligned} \tag{17}$$

We therefore conclude that, as the tissue expands and $X_0(t) \to \infty$, the dynamics is asymptotically dominated by merger and fragmentation processes. Specifically, we can distinguish three asymptotic regimes, which are fixed points of the renormalisation group flow:

1. $\alpha > \gamma$: fragmentation processes dominate fluctuations;
2. $\alpha = \gamma$: merging and fragmentation processes contribute equally to fluctuations;
3. $\alpha < \gamma$: Merging processes dominate fluctuations.

In each of these regimes large-scale fluctuations are dominated by different processes. To continue our analysis, we now study each of these regimes in more detail.

### 1.2 Existence of scaling solutions

We first ask whether scaling solutions, as found empirically in the context of mouse heart development, exist in these asymptotic regimes. Specifically, we are interested in solutions of the form

$$f(x,t) = \psi\left(x/s(t)\right) , \tag{18}$$

where $s(t)$ is a characteristic scale such as, for example, the average cluster size. We first note that solutions that converge to a stationary form under renormalisation, $\vartheta_s(\rho)$, imply scaling solutions in the unrescaled coordinates,

$$f(x,t) = \vartheta_s(x/X_0(t)) . \tag{19}$$

Secondly, scaling solutions under renormalisation lead to scaling solutions in the original coordinates,

$$f(x,t) = \vartheta\left[\rho/\langle\rho\rangle(\tau(t))\right] = \psi\left[x/\langle x\rangle(\tau(t))\right] , \tag{20}$$

where $\tau(t)$ is a rescaled time as defined below. Therefore, the cluster size distribution obtains a scaling form if the renormalised kinetics admit stationary or scaling solutions. In the following we will investigate whether these conditions are fulfilled in the three identified asymptotic regimes.

### 1.2.1 $\alpha = \gamma$: *Merging and fragmentation*

We first consider the regime where merging and fragmentation both contribute to large scale fluctuations, i.e. $\varphi X_0(t)^{\alpha+1} \approx \mu X_0(t)^{\gamma+1}$. We rescale time according to $t \to \int_0^t X_0(t')^{\alpha+1}\mathrm{d}t' \equiv \tau(t)$ and rewrite the time evolution equation as

$$\begin{aligned}\frac{\partial}{\partial\tau}\vartheta(\rho,\tau) \approx \varphi &\left[2\int_0^\infty F_c(\rho,\rho')\vartheta(\rho+\rho',\tau)\mathrm{d}\rho' \right.\\ &\left. -\vartheta(\rho,\tau)\int_0^\rho F_c(\rho-\rho',\rho')\mathrm{d}\rho'\right]\\ +\mu &\left[\int_0^\rho K(\rho',\rho-\rho')\vartheta(\rho,\tau)\vartheta(\rho-\rho',\tau)\mathrm{d}\rho'\right.\\ &\left. -\vartheta(\rho,\tau)\int_0^\infty K(\rho,\rho')\vartheta(\rho',\tau)\mathrm{d}\rho'\right] ,\end{aligned} \tag{21}$$

with time-independent parameters $\varphi$ and $\mu$. After rescaling of cluster sizes and time, the dynamics therefore asymptotically follows a merging-fragmentation dynamics which has been studied in the literature [3].

The coagulation-fragmentation process is comprised of two competing processes: merging increases the average cluster size and decreases the number of clusters, while fragmentation decreases the average cluster size and increases the number of clusters. There exists a crossover time $\tau^*$ after which a the dynamics has converged to a "critical" balance between the two competing processes, and the distribution of cluster sizes, as well as the number of clusters, become independent of time. It has been shown that the fragmentation-merger (viz. coagulation) process converges to a stationary distribution, $\vartheta_s(\rho)$, for a large class of merging and fragmentation kernels [3]. Specifically, merging-fragmentation processes give rise to stationary solutions if $\alpha + 1 > \gamma$ [4]. As we will show below, this is indeed satisfied in the biological context of interest here. The size distribution therefore exponentially converges to a scaling form

$$f(x,t) = \vartheta_s\left(x/X_0(t)\right) \tag{22}$$

on a time scale determined by $\varphi X_0(t)^{\alpha+1}$ or $\mu X_0(t)^{\gamma+1}$.

Here, we should note that we have to be cautious about the limitations of the renormalisation approach. We implicitly assumed that $\langle\rho\rangle$ is of order 1, and, in particular, does not scale inversely with the organ size. This assumption would not be true if the dynamics was stationary in the unrescaled coordinates. In this case we expect typical rescaled cluster sizes to scale inversely with the tissue size, $\langle\rho\rangle \propto X_0(t)^{-1}$. Obviously, stationary solutions do not exist for pure growth-merger processes. For growth combined with merger and fragmentation, we will see that the dynamics is not stationary either. But in the absence of merger, the cluster size distributions is, in many cases, stationary and the growth process might not become irrelevant for large times.

For merging-fragmentation processes, this assumption on typical values of the rescaled cluster size is indeed valid as long as $\alpha + 1 > \gamma$. To see this, we consider the dynamics of the first moment, $\langle\rho\rangle(t) \equiv \int_0^\infty \rho\vartheta(\rho, t)\mathrm{d}\rho$. Its time derivative can be written as the sum of contributions from the growth and merging-fragmentation processes,

$$\partial_t\langle\rho\rangle(t) = \partial_t\langle\rho\rangle_{\text{growth}}(t) + \partial_t\langle\rho\rangle_{\text{frag.+coag}}(t)\,. \tag{23}$$

By the definition of the rescaling, the growth dynamics is neutral in the rescaled coordinates, $\partial_t\langle\rho\rangle_{\text{growth}}(t) = 0$. Furthermore, the contributions from merging and fragmentation processes cancel, such that the typical cluster size in the rescaled coordinates is constant. We conclude that, in the unrescaled coordinates, $\langle x\rangle \propto X_0(t)^\beta$ and the contributions from the growth dynamics to large-scale fluctuations indeed vanish, giving rise to scaling solutions.

### 1.2.2 $\alpha < \gamma$: *Merging*

If merging asymptotically dominates fluctuations, $X_0(t)^{\alpha+1} \ll X_0(t)^{\gamma+1}$ for $t \to \infty$, the time evolution of the cluster size distribution can be written in the form of a coagulation equation. To this end we rescale time, $t \to \int_0^t X_0(t')^{\gamma+1}\mathrm{d}t' \equiv \tau(t)$, and obtain asymptotically

$$\begin{aligned}\frac{\partial}{\partial\tau}\vartheta(\rho,\tau) \approx \mu \Bigg[\int_0^\rho K(\rho',\rho-\rho')\vartheta(r,\tau)\vartheta(\rho-\rho',\tau)\mathrm{d}\rho' \\ -\vartheta(\rho,\tau)\int_0^\infty K(\rho,\rho')\vartheta(\rho',\tau)\mathrm{d}\rho\Bigg]\,,\end{aligned} \tag{24}$$

where $\mu$ is the time-independent merging rate. This equation is known as *Smoluchowski's coagulation equation*, which has been studied in the context of aerosols in the continuum regime among others. While analytical solutions are only known for a few simplistic kernels, it has been shown that, for homogeneous kernels, Equation (24) gives rise to scaling solutions of the form $\vartheta(\rho, \tau) = \psi[\rho/\langle\rho\rangle(\tau)]$ [1]. Consequently, in the context of clonal dynamics in tissue development, we have scaling solutions of the form

$$f(x,t) = \psi\left[\frac{x}{\langle x\rangle(\tau(t))}\right], \tag{25}$$

with $\tau(t) = \int_0^t X_0^{\gamma+1}(t')\mathrm{d}t'$.

### 1.2.3 $\alpha > \gamma$: *Fragmentation*

We finally consider the case where $X_0(t)^{\alpha+1} \gg X_0(t)^{\gamma+1}$ for $t \to \infty$. If $\alpha + 1 > \beta$, then growth and fragmentation give rise to a stationary distribution and scaling behaviour is trivially fulfilled. On the other hand, if $-1 < \alpha + 1 \leq \beta$, typical cluster sizes increase with $X_0(t)$ and we can again asymptotically neglect the growth term. Rescaling time according to $t \to \int_0^t X_0(t')^{\alpha+1}\mathrm{d}t' \equiv \tau(t)$, the cluster size distribution then asymptotes to the form

$$\begin{aligned}\frac{\partial}{\partial\tau}\vartheta(\rho,\tau) \approx \varphi \Bigg[ & 2\int_0^\infty F(\rho,\rho')\vartheta(\rho+\rho',\tau)\mathrm{d}\rho' \\ & - \vartheta(\rho,\tau)\int_0^\rho F(\rho-\rho',\rho')\mathrm{d}\rho'\Bigg].\end{aligned} \tag{26}$$

The fragmentation equation again admits scaling solutions $\vartheta(\rho, \tau) = \psi[\rho/\langle\rho\rangle(\tau)]$ for homogeneous kernels [5], such that the cluster size distribution obtains a scaling form,

$$f(x,t) = \psi\left[\frac{x}{\langle x\rangle(\tau(t))}\right]. \tag{27}$$

This means that, while we find scaling behaviour in both cases, the scaling function is universal only for $-1 < \alpha + 1 \leq \beta$.

In summary, the large-scale behaviour of labelled cell clusters in developing tissues is dominated by large fluctuations stemming from merging and fragmentation processes. The asymptotic behaviour is characterised by one of three possible regimes, which are fixed points of the renormalisation process. Within such a regime, the shape of the cluster size distribution is asymptotically independent of cell fate specific processes and takes a universal scaling form, which is defined by a single characteristic scale.

## 1.3 Numerical simulations

To test these analytical results we performed Monte Carlo simulations both approximating the mean-field kinetics and resembling a simple spatially extended growing tissue.

### 1.3.1 *Monte Carlo simulations of the mean-field Master equations*

To approximately solve the mean-field Master equations we performed kinetic Monte Carlo simulations. Starting from an exponential cluster size distribution at each Monte Carlo cycle, one of the three processes - growth, merger and frag-mentation - was randomly selected with probabilities proportional to the overall rates of these processes,

$$\begin{aligned} &\int_0^\infty \mathcal{L}_{\text{growth}}[f(x,t)]\text{d}x\,, \\ \mu &\int_0^\infty \mathcal{L}_{\text{merging}}[f(x,t)]\text{d}x\,, \text{ and} \\ \varphi &\int_0^\infty \mathcal{L}_{\text{fragmentation}}[f(x,t)]\text{d}x\,, \end{aligned} \tag{28}$$

respectively. For the merging process, the calculation of these rates involves a double integral at each Monte-Carlo cycle, which is numerically unfeasible. We therefore approximated the overall merging rate in the following way: Following Ref. [6] the overall merging rate can be written as

$$\mu/2 \int_0^\infty \int_0^\infty K(x,y) f(y,t) f(x,t) \text{d}x\text{d}y\,. \tag{29}$$

We assume a log-normal cluster size distribution, which is common practice in the literature, and will be justified *post hoc*,

$$f(x,t) = \frac{N(t)}{2\pi\sigma x} \exp\left[-\frac{(\ln x - \mu)^2}{2\sigma^2}\right]\,, \tag{30}$$

with logarithmic mean and standard deviation $\mu$ and $\sigma$, respectively. Substituting this ansatz into the overall merging rate, we can relate the merging rate to the moments of the log-normal cluster size distribution,

$$\int_0^\infty \mathcal{L}_{\text{merging}}[f(x,t)]\text{d}x \approx \mu\left[N(t)^2 + M_{1/d_f} M_{-1/d_f}\right]\,. \tag{31}$$

These moments are related by

$$M_k = N(t) x^k \exp\left(\frac{9}{2} k^2 \ln^2 \sigma\right)\,. \tag{32}$$

For the merging kernel defined in Eq. (42), we therefore find

$$\int_0^\infty \mathcal{L}_{\text{merging}}[f(x,t)]\mathrm{d}x \approx -\mu \left[1 + \exp\left(\frac{9\ln^2\sigma}{d_f^2}\right)\right] N(t)^2 . \tag{33}$$

The right hand side is an approximation to the overall merging rate, which we used to calculate the probability of merging events in each Monte Carlo cycle.

In the next step, clusters or pairs of clusters are randomly drawn from the population according to the statistical weights encoded in the kernels and the reaction is performed. Finally, the simulation time is advanced by the inverse sum of the overall rates,

$$\Delta t = \left\{ \int_0^\infty \mathcal{L}_{\text{growth}}[f(x,t)] + \mu \mathcal{L}_{\text{merging}}[f(x,t)] + \varphi \mathcal{L}_{\text{fragmentation}}[f(x,t)]\mathrm{d}x \right\}^{-1} . \tag{34}$$

Based on the stochastic trajectories obtained from the Monte Carlo simulations, we approximated $f(x,t)$ by calculating histograms. After each Monte Carlo cycle we also calculated the change in variance of the cluster size distribution at a given time, $\Delta\mathrm{var}(x)_t$ and, for all reactions of a given type, calculated the accumulated variance as , $\sum_{t' \leq t} \Delta\mathrm{var}(x)_{t'}$, confirming the scaling of the fluctuations stemming from different processes (Figure S1A). For numerical efficiency we chose in this case constant merging and fragmentation kernels. Parameters were $\varphi = 10$, $\mu = 0.28$ and we averaged over 18 simulation runs with 100 initial clusters each. As a predicted by our calculations, the cluster size distribution converges to a form that is independent of the specific kind of growth dynamics (Figure 2C of the main text). For these simulations we used more realistic kernels as derived below. Parameters were $\varphi = 2e-4$ and $\mu = 2e-3$ for asymmetric divisions, and $\varphi = 1$ and $\mu = 10$ for symmetric divisions. Histograms were taken over 1000 simulations runs involving initially 200 clusters each.

### 1.3.2 *Lattice simulations*

To further test whether the merging and fragmentation, and the ensuing scaling behaviour, arise as an emergent property of the collective dynamics of cells in a growing tissue, we performed lattice simulations of a clonally labelled expanding cell population. While tissue development comprises multiple processes, including collective cell migration and responses to mechanical cues, in order to simulate multiple orders of magnitude in cluster sizes we studied a highly simplified system: cells are arranged on a cubic lattice. Initially, the lattice is of dimension $10 \times 10 \times 10$ cells and each cell (or clone) is assigned a unique identifier. At each time step, points on the dual lattice are randomly occupied by off-spring of neighbouring cells, mimicking the expansion of the tissue. In accordance with our calculations in the long term merging and fragmentation processes dominate contributions to the variance. Coarse-graining was performed using a Gaussian kernel smoother with a standard deviation proportional to the linear lattice size. Again, cluster size distributions collapsed onto scaling forms after few rounds of cell divisions.

# 2 SHAPE OF THE SCALING SOLUTION

Having determined the conditions that lead to the emergence of scaling, we turn now to consider the shape of the scaling solutions. At this point it is important to keep in mind that the sample size resulting from lineage tracing experiments is generally small and, in most cases, it is impossible to distinguish between similar distributions or to resolve behaviour in the tails of the distribution. Further, exact analytical solutions to the fragmentation-merging equations are known only for trivial kernels. For practical purposes, in the context of lineage tracing experiments, we therefore seek to define the simplest approximation to the scaling form within the limits of the experimental context.

To infer the shape of the scaling form we first need to derive the specific functional forms of the merging and fragmentation kernels.

## 2.1 Derivation of the merging kernel

To derive the merging kernel, we note that cluster positions fluctuate due to random mechanical forces from the surrounding tissue. We assume that the characteristic length scale of these fluctuations is much smaller than the average distance between clusters labelled in the same colour, i.e. merging is limited by the motion of clusters in the tissue. We also assume that forces acting on a given cluster are isotropic. In this case, the stochastic motion of clusters is diffusive [7]. We here derive the merging kernel for a three-dimensional system and spherical clusters. However, as discussed further below, the calculations are straightforwardly extendable to other spatial dimensions and non-spherical clusters.

We begin by considering the merging of clusters with a central absorbing cluster with radius $r_0$. Without loss of generality, we can assume that this central cluster is static. The concentration $c(r,t)$ of clusters at a distance $r \geq r_0$ from the central cluster then evolves according to a diffusion equation,

$$\frac{\partial}{\partial t} c(r,t) = \frac{1}{r^2} \frac{\partial}{\partial r} \left[ D r^2 \frac{\partial}{\partial r} c(r,t) \right] , \tag{35}$$

where the diffusion constant $D$ depends on the size of fluctuations and the mechanical properties of the tissue. As we assume that collisions with the central cluster lead to irreversible merging of the clusters, the concentration at its boundary at $r_0$ must vanish. At large distances, $r \to \infty$, cluster concentrations converge to a stationary value, $c_\infty$. The solution of Eq. (35) is therefore given by

$$c(r,t) = c_\infty \left[ 1 - \frac{r_0}{r} \left( 1 - \frac{2}{\sqrt{\pi}} \int_0^{\frac{r - r_0}{2\sqrt{Dt}}} e^{-s^2} \mathrm{d}s \right) \right] . \tag{36}$$

The rate of merging events with the central cluster between times $t$ and $t + dt$ follows as

$$J \, \mathrm{d}t = - \left[ 4\pi r^2 D \frac{\partial}{\partial r} c(r,t) \right]_{r=r_0} \mathrm{d}t \, . \tag{37}$$

With Eq. (36), we obtain for the merging rate

$$J \, \mathrm{d}t = 4\pi D r_0 c_\infty \left( 1 + \frac{r_0}{\sqrt{\pi D t}} \right) \mathrm{d}t \, , \tag{38}$$

and the merging kernel for the central cluster therefore is $K(r_0, t) = J\mathrm{d}t/c_\infty$, or

$$K(r_0, t) = 4\pi D c_\infty \left(1 + \frac{r_0}{\sqrt{\pi D t}}\right) \mathrm{d}t \,. \tag{39}$$

In the asymptotic limit, $t \to \infty$, the dynamics reaches a stationary state and the kernel reduces to $K(r_0) = 4\pi D r_0$. While this result was obtained for a stationary central cluster, it can be straightforwardly extended to obtain the merging kernel of two independent clusters. To this end we consider the relative fluctuations between these two clusters and note that the merging frequency of clusters with radii $r_1$ and $r_2$ is equal to that of a central absorbing cluster with radius $r_1 + r_2$. Similarly, the relative diffusion constant is $D_1 + D_2$. By substituting $r_0 = r_1 + r_2$ and $D = D_1 + D_2$, we finally obtain for the merging kernel

$$K(r, r') = 4\pi(r_1 + r_2)(D_1 + D_2) \,. \tag{40}$$

In three spatial dimensions, the size of a cluster is given by $x = (4/3)\pi r^3$ and the diffusion constant is proportional to the inverse mass, $D \propto x^{-1}$. We finally obtain for the coagulation kernel in three spacial dimensions

$$K(x, x') \propto \left(x^{\frac{1}{3}} + x'^{\frac{1}{3}}\right)\left(x^{-\frac{1}{3}} + x'^{-\frac{1}{3}}\right) \,. \tag{41}$$

This kernel corresponds to the collision kernel used to study aerosols in the continuum regime. While we obtained this result for the special case of three spatial dimensions, it is straightforwardly extendable to other spatial dimensions.

Taking into account the non-spherical structure of clusters, it has been shown that the merging kernel can in general be written as

$$K(x,x') \propto \left(x^{\frac{1}{d_f}} + x'^{\frac{1}{d'_f}}\right)\left(x^{-\frac{1}{d_f}} + x'^{-\frac{1}{d'_f}}\right), \tag{42}$$

where $d_f$ and $d'_f$ are the fractal dimensions of the clusters [8]. For simplicity, here we set $d_f = 2$ noting that our results do not sensitively depend on the specific choice of $d_f$ [Figure 3(a)].

### 2.2 Derivation of the fragmentation kernel

To determine the fragmentation kernel, we write $F(x-x',x') = a(x)b(x'|x)$, where $a(x)$ is the overall rate of fragmentation of clusters of size $x$ and $b(x'|x)$ is the conditional probability of daughter fragment sizes. Homogeneity of the fragmentation kernel implies that the overall fragmentation rate is of the form $a(x) \propto x^{\alpha+1}$ and the distribution of daughter fragments should only depend on the ratio of the size of the daughter fragment to the size of the original cluster, i.e. $b(x'|x) \propto x^{-1}\tilde{b}(x'/x)$ [5]. The overall rate of fragmentation typically increases with the cluster size in most fragmentation processes, such that $\alpha + 1 > 0$. To test if this is the case for the fragmentation of genetically labelled cell clusters in organ development, we analysed recently published clonal data on early heart development [9]. In this work, we employed statistical inference to filter for groups of clonal fragments enabling us to compare the number of fragments of a given clone with its overall size. We indeed found that the number of fragments was proportional to the clone size [Figure 2(a)]. In the following, we therefore set $\alpha = 0$ such that the overall rate of fragmentation increases linearly with the cluster size. For the conditional distribution of daughter fragments, $\tilde{b}(x'/x)$, we could assume a uniform distribution, $\tilde{b} \equiv 1$. However, this would in principle lead to infinitesimally small fragment sizes and an unrealistic power law tail in the distribution for small fragments after repeated rounds of fragmentation [7]. However, the sizes of daughter fragments are effectively limited for two reasons:

1. The coarse-graining procedure imposes a lower limit on the possible sizes of daughter fragments. This limit is of the order $\langle \rho X_0(t) \rangle$. If typical cluster sizes are large, the resolution of the microscope limits the quantification of small fragments. Even above the detection threshold, if small fragments are in the vicinity of a larger cluster, their respective sizes are typically combined in the quantification process.

2. If typical cluster sizes are small, the sizes of daughter fragments cannot be smaller than single cells.

Realistic fragmentation kernels therefore cannot produce infinitesimally small fragments. We take into account this fact by introducing a cut-off to the uniform fragment distribution, i.e. $\tilde{b}(z) = z_c \theta(z - z_c)$, where $z = x/x'$ and $z_c$ is the cut-off. It is important to note that the specific shape of the fragment distribution does not sensitively alter the results as long as the production of very small fragments is limited. To test this assumption, we may again refer to the reconstructed clones from the clonal assay. We can estimate $\tilde{b}(z)$ by calculating the distribution of fragment sizes divided by the overall clone size for clones consisting of two fragments. We find that $\tilde{b}(z)$ indeed vanishes for small fragment sizes and, equivalently, is peaked a $1/2$ [Figure 2(b)]. In principle, this is also consistent with a scenario, where fragmentation only occurs early after induction. In this case, the distribution of relative fragment sizes in clones consisting of three fragments would be peaked at 1/3, which is not the case [Figure 2(c)].

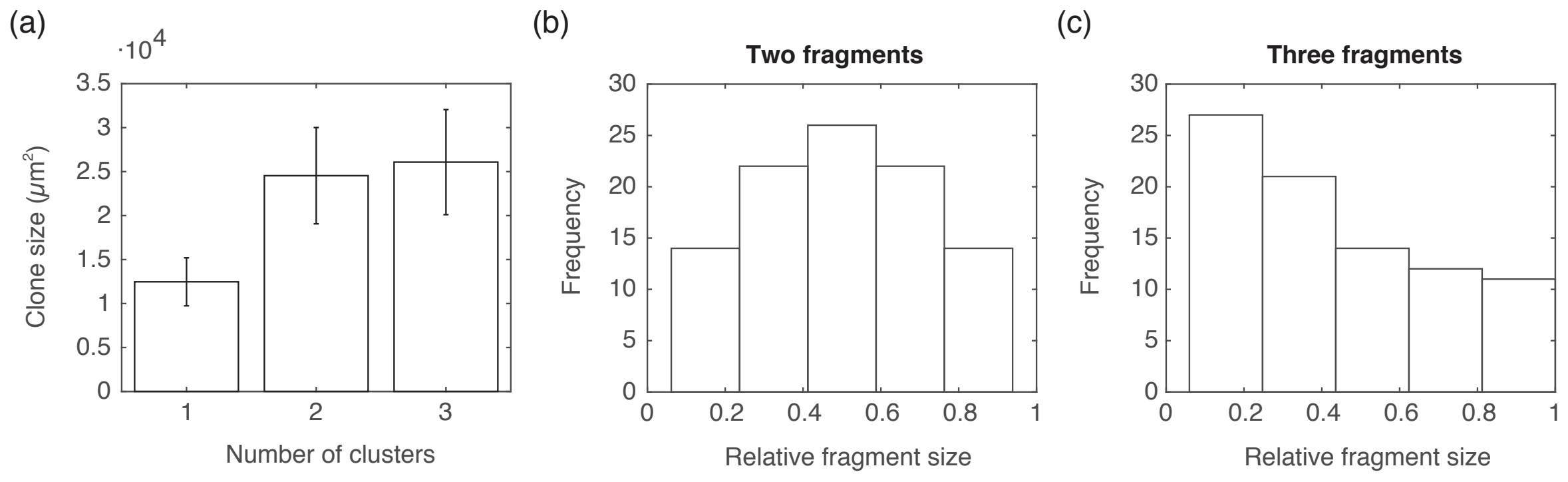

**Figure 2:** Statistics of the sizes of monoclonal fragments (a) Average clone size as a function of the number of fragments. Error bars signify 95% confidence intervals. (b) Histogram of the relative sizes of fragments (proportion of the total clone size) in clones consisting of two fragments. (b) Histogram of the relative sizes of fragments in clones consisting of three fragments.

We emphasize that the specific shape of $\tilde{b}(z)$ does not sensitively alter the shape of the scaling form, as long as fragments of very small size are unlikely. Indeed, we can rewrite the fragmentation terms as

$$\mathcal{L}_{\text{fragmentation}}[f(x,t)] = \int_0^\infty \frac{a(x+x')}{x+x'} \tilde{b}\left(\frac{x'}{x+x'}\right) f(x+x',t)\mathrm{d}x' - a(x)f(x,t)\,. \quad (43)$$

This shows that, for $x \to \infty$, the fragmentation dynamics is dominated by the loss term, which only depends on $a(x)$. Therefore, the right tail of the cluster size distribution may depend on the overall rate of fragmentation, $a(x)$, but it does not depend sensitively on the conditional distribution of daughter fragments. In fact, in lineage tracing studies in developing tissues, the left tail of the cluster size distribution is often not measurable. To summarize, the fragmentation kernel takes the form

$$F_c(x-x',x') = \frac{a(x)}{x}\tilde{b}\left(\frac{x'}{x}\right) = z_c\theta\left(\frac{x}{x'} - z_c\right)\,, \quad (44)$$

which corresponds to uniform fragmentation at an overall rate proportional to the cluster size.

## 2.3 Shape of the scaling function

Having defined the merging and fragmentation kernels, we are now in a position to calculate the shape of the scaling form. Both kernels are homogeneous with exponents $\gamma = \alpha = 0$. Hence, asymptotically, both processes contribute to the scaling form and the rate of convergence depends on the rates $\varphi$ and $\mu$. In other words, large scale fluctuations are controlled by a renormalisation group fixed point corresponding to a critical state dominated by a balance between merging and fragmentation.

To begin we consider the regime where large scale fluctuations are dominated by both merging and fragmentation processes. To calculate the scaling form we approximated the solution of Eq. (21) with the kernels defined in (42) and (44) by Monte Carlo simulations. We found that the dynamics reaches a stationary state which is well described by a log-normal cluster size distribution,

$$\frac{f(x,t)}{N(t)} = \frac{1}{2\pi\sigma x} \exp\left[-\frac{(\ln x - \mu)^2}{2\sigma^2}\right], \tag{45}$$

where $N(t)$ is the total number of clusters at time $t$ post-labelling and $\mu$ and $\sigma$ are the mean and standard deviation of $\ln x$, respectively (Figure 3(b)). Small deviations from the log-normal shape in the tail corresponding to small cluster sizes were found if the cutoff is small. In the extreme case of a vanishing cutoff the left tail decays algebraically.

In the case that $\gamma > \alpha$, merging dynamics dominate the large-scale behaviour. Eq. (24) has been studied extensively by analytical and numerical approximations. It has been found that the merging dynamics gives rise to scaling solutions, whose shape is well approximated by a log-normal distribution [1]. Similarly, a log-normal distribution is also found if fragmentation dominates the large-scale dynamics [5]. Importantly, the occurrence of log-normally distributed cluster sizes in all three cases is a result of the coarse-graining procedure, which induces a small size cutoff in the fragmentation kernel. Values for the standard deviation of loga-rithmic cluster sizes obtain values between 0.5 and 1.5 depending on the precise values of the cutoff and the degree of homogeneity of the fragmentation kernel and compared to a fitted value of 1.05 obtained for the universal curve plotted in Fig. 3F of the main text.

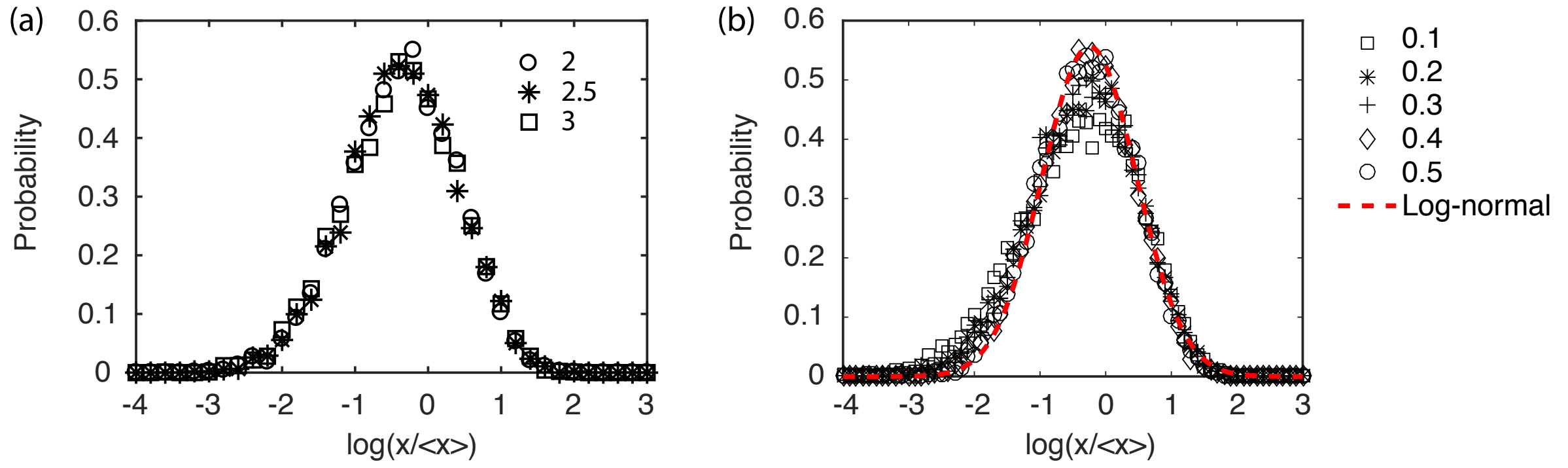

**Figure 3:** (a) Asymptotic solution of the merging-fragmentation equation (21) for different values of the (fractal) dimension of labelled clusters, $d_f$. The shape of the scaling form is independent of $d_f$. (b) Asymptotic solution the merging-fragmentation-fragmentation equation with $d_f = 2$, and different values of the fragmentation cutoff. The probability distributions agrees excellently with a log-normal form. For small values of the cutoff a slight deviation from the log-normal form is observed in the left tail, which approaches a power-law decay for vanishing cutoff. Monte Carlo simulations were started with an initial exponential distribution of cluster sizes and 200 clusters with $\mu = 10^{-5}$ and $\varphi = 1$. Each probability distribution was calculated at $\tau = 5$, such that each cluster, on average, underwent roughly 50 merging and fragmentation events. Histograms were calculated by pooling 1000 runs for a given set of parameters.

In the absence of such a cutoff, i.e. if fragment sizes were observable down to the smallest scale, cluster size distributions follow slightly different dependencies. While to our knowledge no exact solution of Eq. (21) with the merging and fragmentation kernels defined in Eqs. (42) and (44), respectively, is known we can hope to gain analytical insight by studying merging-fragmentation dynamics of a simpler system which comprises essential features of the dynamics. Such a simplification can be obtained by choosing a constant merging kernel which has the same degree of homogeneity as the one defined in Eq. (42). With this choice, Eq. (42) simplifies to

$$\begin{aligned}\frac{\partial}{\partial\tau}\vartheta(\rho,\tau) \approx \varphi\left[2\int_x^\infty \vartheta(\rho',\tau)\mathrm{d}\rho' - \rho\vartheta(\rho,\tau)\right] \\ + \mu\left[\int_0^\rho \vartheta(\rho,\tau)\vartheta(\rho-\rho',\tau)\mathrm{d}\rho' - \vartheta(\rho,\tau)\int_0^\infty \vartheta(\rho',\tau)\mathrm{d}\rho'\right].\end{aligned} \tag{46}$$

Following Ref. [10], a solution can be obtained by studying the time evolution of the Laplace transform of $\vartheta(\rho,\tau)$,

$$L(\sigma,\tau) = \int_0^\infty e^{-\sigma\rho}\vartheta(\rho,\tau)\mathrm{d}\rho\,. \tag{47}$$

Applied to Eq. (46) we obtain condition for the stationarity of the solution,

$$\frac{\partial L(\sigma,\tau)}{\partial\sigma} = -\frac{2}{\sigma} + \frac{2L(\sigma,\tau)}{\sigma} + \frac{2L(\sigma,\tau)}{\sqrt{\rho_0\varphi/(N(0)\mu)}} - \frac{N(0)\mu L(\sigma,\tau)^2}{\rho_0\varphi}, \tag{48}$$

with the total rescaled mass of clusters $\rho_0 = \int_0^\infty \rho\vartheta(\rho,\tau)\mathrm{d}\rho$ and the total initial number of clusters, $N_0$. The solution is

$$L(\sigma,\tau) = \frac{N(0)\mu}{\rho_0\varphi}\left(\sqrt{\frac{N(0)\mu}{\rho_0\varphi}} + \sigma\right)^{-1}. \tag{49}$$

Inverting the Laplace transform finally yields the stationary solution

$$\vartheta_s(\rho) = \frac{N(0)\mu}{\rho_0\varphi}e^{-\sqrt{\frac{N(0)\mu}{\rho_0\varphi}}\rho}, \tag{50}$$

and, consequently, the cluster size distribution follows as

$$f(x,t) \approx \vartheta_s(x/X_0(t)) = \frac{N(0)\mu}{\rho_0\varphi}e^{-\sqrt{\frac{N(0)\mu}{\rho_0\varphi}}\frac{x}{X_0(t)}}. \tag{51}$$

Finally, the cluster size probability distribution is obtained by dividing by the number of clusters in steady state, $N(t) = \sqrt{N(0)\mu/(\rho_0\varphi)}$, such that

$$p(x,t) \approx \sqrt{\frac{N(0)\mu}{\rho_0\varphi}} X_0(t)^{-1} e^{-\sqrt{\frac{N(0)\mu}{\rho_0\varphi}}\frac{x}{X_0(t)}} . \tag{52}$$

The size distribution of labelled clusters therefore follows an exponential form. The exponential size dependence is demonstrated by numerical simulations (Fig. 4), where we also took into account a scenario where merging and fragmentation can occur during turnover of an adult tissue (homeostasis). In this case the time evolution of $f(x,t)$ is described by

$$\frac{\partial}{\partial t} f(x,t) = \frac{\partial^2}{\partial x^2} f(x,t) + \varphi \mathcal{L}_{\text{fragmentation}}[f(x,t) + \mu \mathcal{L}_{\text{merging}}[f(x,t)] , \tag{53}$$

with an absorbing boundary condition at $x = 0$. The diffusion term describes neutral dynamics of cluster sizes due to loss and replacement of stem cells and similar to the growth terms it becomes irrelevant under the renormalisation group transformation. Asymptotically, universality holds down to the smallest scales resolvable by our simulations. For small cluster sizes we found a deviation from the exponential form. While the exact functional form of the small size dependence cannot be unambiguously inferred from the simulations, our numerical results are in agreement with an algebraic dependence of the scaling form for small cluster sizes, $x^\alpha$ with $\alpha \approx 1/3$ and an exponential decay for $x \gg \langle x \rangle$.

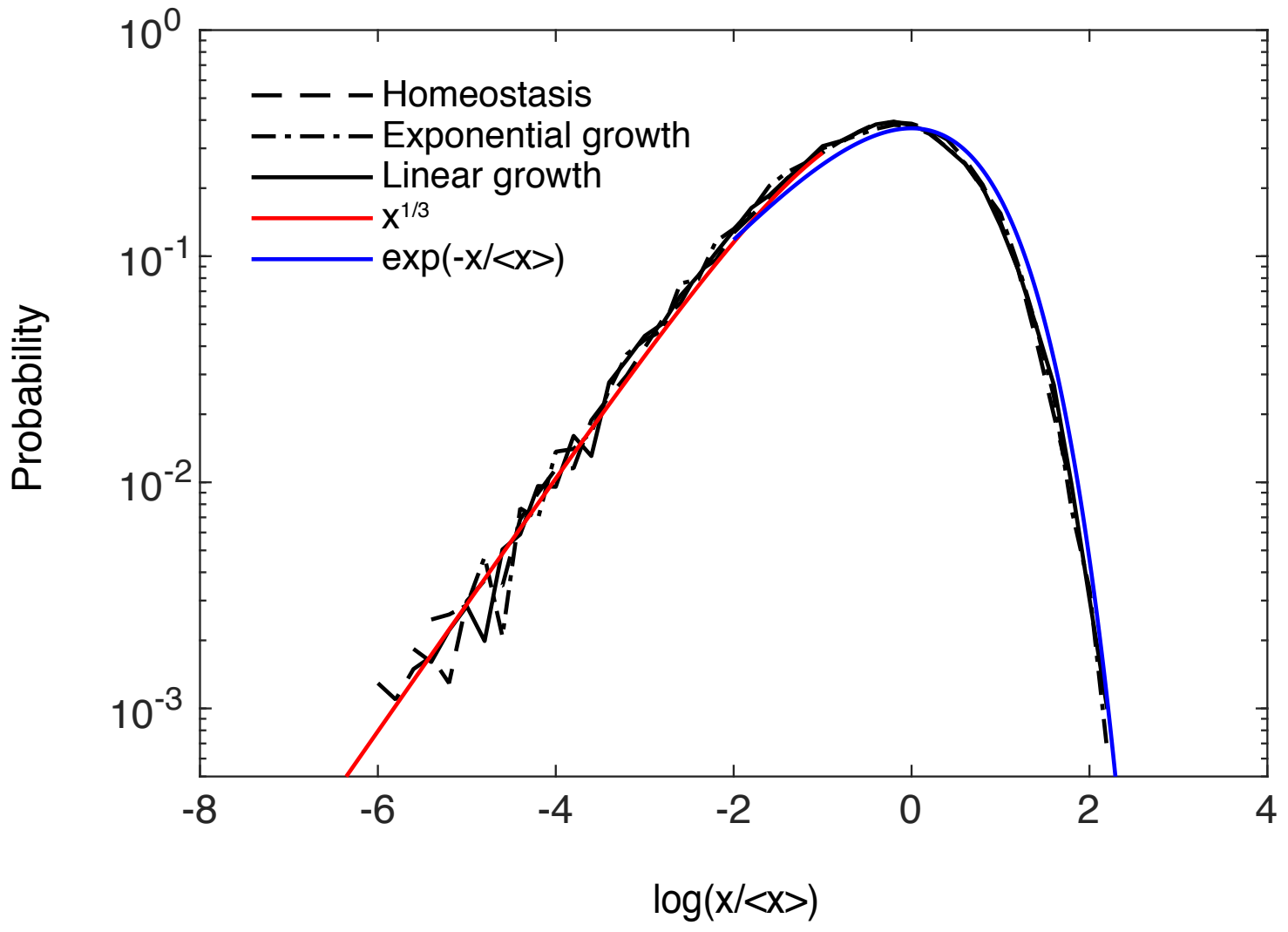

**Figure 4:** Asymptotic numerical solutions of the time evolution equations (4). Solutions show scaling behaviour and universality. Fits for small and large cluster sizes are coloured red and blue, respectively. Parameters were chosen such that $\langle x \rangle \gg 1$: Homeostasis: $\varphi = 5 \cdot 10^{-4}$, $\mu = 5 \cdot 10^{-9}$; linear growth: $\varphi = 5 \cdot 10^{-5}$, $\mu = 1 \cdot 10^{-10}$; exponential growth: $\varphi = 0.25$, $\mu = 2.5 \cdot 10^{-6}$. Histograms were calculated by pooling 1000-2000 runs for a given set of parameters.

# 3 NON-UNIVERSAL DEPENDENCIES OF THE CLUSTER SIZE DISTRIBUTION

The emergence of scaling behaviour and the universality of the scaling functions gives rise to challenges in the interpretation of lineage tracing experiments in developing tissues. The size distributions of labelled clusters are asymptotically independent of the details of the biological context. How can lineage specific infor-mation then be retained in genetic tracing experiments? Our understanding of the emergence of universal scaling behaviour in fact allows for the identification of strategies to reveal cell fate behaviour. Specifically, non-universal dependences can be recovered in several ways:

1. Convergence to scaling behaviour occurs exponentially on a time scale determined by $\mu$ and $\varphi$. Cell fate specific information can therefore be retained from the short term dynamics, where time is much shorter than the time scales associated with the merging and fragmentation rates, as measured in units of the cell cycle time.
2. Small-size dependencies, $1 \ll x \ll \langle x \rangle$, converge to the universal form last. Importantly, to compare experimental data with the modelling predictions in addition to cell fate related processes, merging and fragmentation needs to be specifically taken into account.
3. Last, while explicit information on cell fate is erased, merging and fragmentation are emergent processes resulting from many cell fate decisions. Given a large enough sample size, the shape of the scaling function might show specific dependences in different kinds of tissues.

Lineage-specific information can be retained from non-universal dependencies at short times and small scales. To this end data from functional assays must be compared to models describing cell fate processes as well as merging and fragmentation dynamics. We here consider situations, where merging is negligible compared to fragmentation, for example due to a sufficiently low induction fre-quency. Then, for linear fragmentation with uniformly distributed fragment sizes, the mean-field Master equation reads

$$\frac{\partial}{\partial t} f(x,t) = -\frac{\partial}{\partial x} x^{\beta} f(x,t) + \varphi \left[ 2 \int_x^{\infty} f(x',t) \mathrm{d}x' - x f(x,t) \right] , \tag{54}$$

The probability distribution of fragment sizes is given by the number of fragments of a given size divided by the total number of fragments in the system, $p(x,t) \equiv f(x,t)/N(t)$. To derive the time evolution equation of $p(x,t)$ we divide Eq. (56) by $N(t)$ and obtain

$$\frac{\partial}{\partial t} p(x,t) + \frac{p(x,t)\dot{N}(t)}{N(t)} = -\frac{\partial}{\partial x} x^{\beta} p(x,t) + \varphi \left[ 2 \int_x^{\infty} p(x',t) \mathrm{d}x' - x p(x,t) \right] . \tag{55}$$

With $\dot{N}(t) = \varphi \int_0^{\infty} x p(x,t) \mathrm{d}x N(t)$, this then yields

$$\begin{aligned} \frac{\partial}{\partial t} p(x,t) = &- \frac{\partial}{\partial x} x^{\beta} p(x,t) + \varphi \left[ 2 \int_x^{\infty} p(x',t) \mathrm{d}x' - x p(x,t) \right] \\ &- \varphi p(x,t) \int_0^{\infty} x p(x,t) \mathrm{d}x \, . \end{aligned} \tag{56}$$

This equation gives rise to a stationary solution, where growth and fragmentation balance. To obtain this solution we differentiate with respect to $x$, such that

$$\frac{\partial}{\partial x^2} x^{\beta} p(x,t) + 2\varphi p(x,t) - \varphi \frac{\partial}{\partial x} x p(x,t) + \varphi \int_0^{\infty} x p(x,t) \mathrm{d}x \, \frac{\partial}{\partial x} p(x,t) = 0 \, . \tag{57}$$

The steady state value of the first raw moment is

$$\int_0^{\infty} x p(x,t) \mathrm{d}x = \varphi^{-\frac{1}{2-\beta}} \, . \tag{58}$$

With this, requiring positivity and normalisation of $p(x,t)$, we find solutions of the form

$$p(x,t) = \varphi x \left(2 + \sqrt{\varphi} x\right) e^{-\sqrt{\varphi} x - \frac{\varphi}{2} x^2} \text{ for } \beta = 0 \, , \tag{59}$$

and

$$p(x,t) = \varphi e^{-\varphi x} \text{ for } \beta = 1\,. \tag{60}$$

If merging is negligible, cluster size distributions for different division modes are therefore distinguishable in their small size limit, where asymmetrically dividing populations give rise to an algebraically increasing size distribution while symmetrically dividing populations are characterised by an exponentially decreasing distribution. It is important to note, however, that this is only the case if there is no small-size cutoff in the fragmentation kernel, i.e. for $1 \ll x \ll \langle x \rangle$. If the typical cluster size is not much larger than the size of a single cell, the effective cutoff in fragment sizes gives rise to a log-normal distribution in both cases.

In summary, by coarse graining the kinetic equations we found that merging and fragmentation of labelled clusters in lineage tracing experiments in developing tissues give rise to universal size distributions. Cell fate specific information is ultimately erased in these experiments. By understanding the origin of this universality our approach allows identifying strategies for retaining cell fate specific information from these experiments.

# A RESCALING OF OTHER CELL FATE PROCESSES

In this appendix we perform the rescaling for cell death and immigration processes. In the framework of the mean-field Master equation these processes contribute terms of the forms

$$\begin{aligned}\text{Cell death: } \partial_t F\big|_{\text{death}} &= \delta\partial_x[xf(x,t)]\,, \\ \text{Immigration: } \partial_t F\big|_{\text{immigration}} &= I\delta(x-x_0)\,.\end{aligned} \tag{61}$$

Although cell death is rare in expanding population an equivalent process can effectively occur in situations, where only a two-dimensional surface of patches is measurable. However, within the mathematical framework of the mean-field Master equation cell death introduces a term of the form $\partial_x xf(x,t)$ on the right hand side, which is the negative analogue to the growth term. Following the calculations on the growth terms we therefore find that the contributions of cell death to the variance become negligible as organ growth proceeds.

Nucleation of new cluster (immigration) of size $x_0$ can arise as a result of continuous induction or, again, as a result of cell migration in sectional data. In rescaled coordinates the immigration term reads $I = I'\delta(\rho - \rho_0)$ with $\rho_0 = \rho/X_0(t)$ and the rescaled rate

$$I' = X_0(t)^{-1} I\,. \tag{62}$$

We therefore find that nucleation of new clusters becomes asymptotically irrelevant.